\documentclass[a4paper,10pt]{article}
\usepackage{graphicx} 
\usepackage[utf8]{inputenc}
\usepackage[T1]{fontenc}
\usepackage{comment}
\usepackage{float}
\usepackage{natbib}
\usepackage{pdflscape}  
\usepackage{xcolor}
\usepackage{booktabs} 
\usepackage{lipsum} 

\usepackage{a4wide}
\usepackage{indentfirst}
\usepackage{authblk}

\usepackage{amsmath,amssymb}
\usepackage{amsthm}
\usepackage[colorlinks=true, linkcolor=black, citecolor=black, urlcolor=blue]{hyperref}
\usepackage{footmisc} 
\usepackage{subfig}
\usepackage{graphicx}
\usepackage{algorithm2e}
\usepackage{subcaption}
\usepackage{chngcntr}
\counterwithin{table}{section}
\counterwithin{figure}{section}

\usepackage[margin=1.2in]{geometry}
\usepackage{graphicx}
\usepackage{listings}
\usepackage[utf8]{inputenc}
\usepackage{appendix}
\usepackage{shuffle}
\numberwithin{equation}{section}
\newtheorem{teo}{Theorem}[section]
\newtheorem*{teo*}{Theorem}
\newtheorem*{prop*}{Proposition}
\newtheorem*{corol*}{Corollary}
\newtheorem{prop}[teo]{Proposition}

\newtheorem{defi}[teo]{Definition}

\theoremstyle{definition}

\newtheorem*{remark*}{Remark}

\newcommand{\floor}[1]{\left\lfloor #1 \right\rfloor}

\newcommand{\bfa}{\mathbf{a}}
\newcommand{\bfb}{\mathbf{b}}

\newcommand{\N}{\mathbb{N}}

\newcommand{\R}{\mathbb{R}}

\newcommand{\bX}{\mathbf{X}}
\newcommand{\bY}{\mathbf{Y}}

\newcommand{\eps}{\varepsilon}

\newcommand{\blfootnote}[2]{%
  \begingroup
  \renewcommand\thefootnote{#1}%
  \addtocounter{footnote}{1}%
  \footnotetext{#2}%
  \addtocounter{footnote}{-1}%
  \endgroup
}

\title{Signature-based identification of volatility models from path geometry}

\author{Òscar Burés\textsuperscript{§} and Rafael {D}e Santiago\textsuperscript{†}}
\date{\today}

\begin{document}

\maketitle

\begingroup
\renewcommand{\thefootnote}{}
\renewcommand{\footnotemargin}{0pt}
\footnotetext{The authors thank Elisa Al{\`o}s, Miguel A. Canela and Josep Vives for carefully reading an earlier version of this manuscript and for their valuable comments and suggestions.}
\endgroup

\blfootnote{§}{Departament de Matemàtica Econòmica, Financera i Actuarial, Universitat de Barcelona. Diagonal 690--696, 08034 Barcelona, Spain.}
\blfootnote{†}{Department of Managerial Decision Sciences, IESE Business School. Av. Pearson 21, 08034 Barcelona, Spain.}

\begingroup
\renewcommand{\thefootnote}{}  
\renewcommand{\footnotemargin}{0pt}  
\footnotetext{%
\noindent\hspace{0pt}Òscar Burés supported by program AGAUR-FI ajuts (2025 FI-1 00580) from the Department of Research and Universities of the Government of Catalonia and the co-funding of the European Social Fund Plus (ESF+).}
\endgroup

\begin{abstract}
We propose a signature-based framework for the identification of stochastic volatility model classes from observed path data. By mapping volatility trajectories into a feature space via truncated path signatures and applying a gradient boosting classifier, we show that it is possible to distinguish between different classes of volatility dynamics without relying on parametric calibration.

Through a series of numerical experiments, we demonstrate that the method achieves high classification accuracy across a range of settings, from structurally distinct models to cases involving rough volatility models with closely spaced Hurst parameters. We show that the method remains effective under parameter uncertainty, where each simulated path is drawn with randomly sampled model parameters, and provide a detailed analysis of the misclassification pattern between the Heston and Ornstein--Uhlenbeck models in terms of the volatility of volatility parameter. The results highlight that most of the relevant discriminative information is captured by the first four levels of the signature, while higher-order terms provide only marginal improvements. Overall, the findings support the view that stochastic volatility models can be effectively identified through the geometry of their sample paths.
\end{abstract}

\noindent\textbf{Keywords:} Path signatures, stochastic volatility models, model identification, gradient boosting, rough volatility, Hurst parameter.

\medskip
\noindent\textbf{JEL Classification:} C55, C58, C63, G13.

\noindent \textbf{MSC 2020:}: 60G22, 60H99, 68T05, 91G20.

\section{Introduction}

A common problem in applied science is how to infer from observed data the \emph{mechanism} underlying a dynamical system. In chemistry, for instance, one typically observes only a series of concentrations over time, while the true mechanism (a sequence of chemical reactions) is hidden. 

Scientists typically extract features from data (such as reaction rates and how these rates depend on concentrations), derive mathematical models (rate laws), and then compare alternative mechanisms manually by calibrating the models and assessing their goodness of fit. 

An alternative approach, made available by the development of machine learning techniques, is to define a finite set of possible mechanisms (classes), and then frame the calibration problem as a \emph{classification} task. Rather than estimating the parameters of a fixed volatility model, one identifies the \emph{type of system} that generated the data.

\cite{BuresLarrosa23} use this approach to identify the reaction mechanism from kinetic data. They simulate a range of candidate mechanisms and train a neural network to classify the underlying mechanism from a small number of observed concentration profiles. The model learns directly from the \emph{shape of the trajectories} and is able to identify the correct mechanism with high accuracy.

Inspired by this approach, we aim to infer the underlying volatility model directly from observed (simulated) paths. Our methodology differs from \cite{BuresLarrosa23} in that, rather than learning from the raw trajectories, we learn from their signature representation. This distinction also influences our choice of learning algorithm and the overall framework. 

In particular, we investigate whether the signature of a volatility path (to be defined precisely below) contains enough information to identify the model driving the volatility dynamics. More precisely, we address the following questions: (i) how accurately can signatures distinguish between Heston, Ornstein-Uhlenbeck and rough Bergomi volatility; (ii) how well can they distinguish between rough Bergomi volatilities generated with different values of the Hurst parameter $H$.

The signature of a path, originally introduced in \cite{chen58} and developed rigorously in \cite{lyons98}, provides a rich and systematic encoding of its temporal structure. By mapping paths into a collection of iterated integrals, signatures capture nonlinear interactions between increments without relying on a specific parametric model. This makes them particularly well suited for representing and distinguishing the behaviors exhibited by different stochastic volatility models.

On the one hand, signatures have been used for the calibration of the implied volatility surface, as in \cite{cuchiero25} and \cite{ABSV26}. In that setting, the volatility is represented as a linear functional of the truncated signature of a driving stochastic process, and the goal is to determine the coefficients of this functional so as to minimize the discrepancy between model-generated asset prices and market asset prices.

On the other hand, a growing body of work has explored the use of machine learning techniques for the calibration of stochastic volatility models, in particular in the rough volatility setting; see, for instance, \cite{stone20}, \cite{horvath21}, \cite{bayer19deep}, \cite{hernandez17} or \cite{kondratyev18}. 

Our objective here is of a different nature. Rather than calibrating model parameters to market data, we address a \emph{model identification} problem: given a volatility trajectory, can we infer the volatility model that generated it? 

\cite{stone20} proposes a convolutional neural network approach to estimate the H\"{o}lder exponent (equivalently, the Hurst parameter $H$) from simulated log-volatility paths of the rough Bergomi model. Calibration is framed as a regression problem, where the network is trained on paths with known $H$ and then used to predict the Hurst parameter from unseen trajectories. Our approach differs in two important respects, which can be summarized as follows:
\begin{center}
\begin{tabular}{lcl}
\textsc{Input}  & & \textsc{Output} \\ [4pt]
Raw path        & $\longrightarrow$ & Parameter estimation (e.g.\ $\hat{H}$) \\ [2pt]
Signature       & $\longrightarrow$ & Model identification (e.g.\ Heston vs.\ Ornstein--Uhlenbeck vs.\ rBergomi) \\
\end{tabular}
\end{center}

\noindent First, rather than estimating a single parameter of a fixed model, we ask which class of stochastic volatility model generated the observed trajectory. Second, instead of feeding raw paths directly into a neural network, we map each path to its truncated signature before classification, providing a structured and theoretically grounded feature representation that captures the full geometry of the path.

In this paper we simulate volatility paths from several benchmark stochastic volatility models, compute their truncated signatures, and train an XGBoost classifier to infer the underlying model class. We consider two settings of increasing difficulty: a controlled setting in which each model class is simulated with fixed parameter values, and a more demanding setting in which parameters are sampled randomly within given ranges, so that the classifier learns to identify model structure rather than specific parameter configurations. Performance is evaluated on out-of-sample trajectories using confusion matrices and classification accuracy.

The remainder of the paper is organized as follows. In Section~\ref{sec:signatures} we introduce the basic elements of rough path theory required to define and compute path signatures. In Section~\ref{sec:vol_models} we describe the volatility models under consideration and discuss how their associated paths can be lifted and represented via signatures. In Section~\ref{sec:xgboost} we present the learning framework, with particular emphasis on the XGBoost algorithm. In Section~\ref{sec:numerical_fixed}, we address the question: ``Can signatures distinguish between path geometries?'' To this end, we perform a series of proof-of-concept experiments in which trajectories within each class are generated using fixed parameter values. 

In Section~\ref{sec:numerical_random}, we turn to the more challenging problem of \emph{model-class identification} under parameter uncertainty, addressing the question: ``Can signatures identify the underlying model class despite changing parameters?'' In this setting, model parameters are randomly sampled for each trajectory, so that the classifier must learn to recognize structural properties of the volatility dynamics rather than features associated with a particular parameter configuration. We conclude in Section~\ref{sec:conclusion} with a discussion of the main findings and possible directions for future research.

\section{Signatures of Rough Paths} \label{sec:signatures}

We introduce here the essential ideas from rough path theory that underpin the signature-based approach. An insightful exposition is given in the Saint-Flour lecture notes by \cite{lyonscaruanalevy07}; see also \cite{lyons02}, \cite{chevyrev16}, and \cite{geng21}. In the paragraphs below we present the main results required for the implementation of signature-based models, omitting proofs. We follow \cite{ABSV26}, which provides a self-contained introduction to the necessary concepts and includes the proofs of most results.

We begin with sufficiently regular trajectories. Let $X:[0,T]\to V$ be a path of bounded variation. Its \emph{signature} is defined as the formal series of iterated integrals:
\[
S(X)_{s,t} = \Big(1,\ \int_s^t dX,\ \int_{s<u_1<u_2<t} dX_{u_1}\otimes dX_{u_2},\ \int_{s<u_1<u_2<u_3<t} dX_{u_1}\otimes dX_{u_2}\otimes dX_{u_3},\ \dots\Big).
\]
Let $X$ take values in $\R^d$. As the $N$-th iterated integral is a tensor with $d^N$ components, the natural framework to organize these objects is that of tensor algebras. We therefore introduce the following concepts. 

Let $V$ be a finite-dimensional real vector space. The \emph{extended tensor algebra} over $V$ is defined as the space of formal series of tensors of all orders:
\[
T((V)) = \{a=(a_0,a_1,\dots): a_n\in V^{\otimes n}\};
\]
The tensor algebra $T(V)$ is defined as the subspace of $T((V))$ consisting of elements with finitely many non-zero terms. The \emph{truncated tensor algebra} of order $N$ is defined as
\[
T^N(V) = \{a \in T((V)) : a_k = 0 \text{ for all } k > N\}.
\]

For any $\bfa, \bfb\in T^N(V)$, we define the truncated tensor product in $T^N(V)$ as
\[
\bfa \otimes_{\leq N}\bfb = \pi_{\leq N}(\bfa\otimes\bfb),
\]
where the projection map $\pi_{\leq N}: T((V))\to T^N(V)$ is defined as $\pi_{\leq N}((a_i)_{i=0}^{\infty}) := (a_i)_{i=0}^N$. If there is no risk of confusion, we will use $\otimes$ in $T^N(V)$ to denote $\otimes_{\leq N}$.

Let $(e_1,\dots,e_d)$ be a basis of $V$, and let $I = (i_1,\dots,i_n) \in \{1,\dots,d\}^n$ be a word of length $n$ over the alphabet $\{1,\dots,d\}$. Then the family
\[
\{ e_I := e_{i_1}\otimes \cdots \otimes e_{i_n} : |I| = n \}
\]
forms a basis of $V^{\otimes n}$.

Given $\ell \in T(V)$ and $a \in T((V))$, with expansions
\[
\ell = \sum_{|I|\ge 0}\ell_I e_I, \qquad a = \sum_{|I|\ge 0} a_I e_I,
\]
we introduce the linear functional
\[
\langle \ell, a \rangle := \sum_{|I|\ge 0} \ell_I a_I,
\]
which is well-defined since $\ell$ has finitely many non-zero coefficients. Note that the coordinates of $a$ can be recovered via $\langle e_I, a \rangle = a_I$. 

Signatures are characterized by the following two important properties. 

\begin{teo}[Chen's identity]
    Let $X: [0,T] \to V$ be a continuous path of bounded variation. Then, for all $t\in (0, T)$,
    \begin{equation} \label{chenidentity}
    S(X)_{0,T} = S(X)_{0,t} \otimes S(X)_{t, T}.
    \end{equation}
\end{teo}

\noindent Chen's identity enables the reconstruction of the signature of $X$ over the interval $[0, T]$ provided that it is known on a collection of subintervals that cover $[0, T]$. 

Signatures also satisfy the \emph{group-like property}. An element $\mathbf{a} \in T((V))$ is said to be group-like if for every pair $\ell^1, \ell^2 \in T(V)$ we have
\[
\langle \ell^1, \bfa \rangle \langle \ell^2, \bfa \rangle = \langle \ell^1 \shuffle \ell^2, \bfa \rangle,
\]
where the shuffle product $\shuffle$ represents all the ways in which two tensors can be combined while preserving their internal order. We denote by $G(V)$ the set of group-like elements of $T((V))$.

\begin{prop}\label{grouplikeprop}
    Let $X:[0,T] \to V$ be a continuous path of bounded variation. Then, the signature of $X$ satisfies the group-like property. That is, for every pair $\ell^1, \ell^2 \in T(V)$,  
    \[
    \langle \ell^1, S(X)_{0,T} \rangle \langle \ell^2, S(X)_{0,T} \rangle = \langle \ell^1 \shuffle \ell^2, S(X)_{0,T} \rangle.
    \]
\end{prop}

\noindent The group-like property says that evaluating the shuffle is equivalent to evaluating each tensor separately and multiplying the results. As a consequence, products of signature coordinates can be expressed as linear combinations of higher-order coordinates.

Let $X: [0,T] \to V$ be a continuous path of bounded variation and $\Delta_{T} = \{(s,t) \in [0,T]^2 ; s \leq t\}$. The truncated signature of order $N$ of a path $X$ can be defined as 
\begin{align*}
S(X)^{\leq N}: \Delta_T & \to T^N(V) \\
(s, t) & \mapsto \pi_{\leq N}(S(X)_{s,t}).
\end{align*}
We write $S(X)^{\leq N}(s,t) = S(X)^{\leq N}_{s,t}$. It follows from Chen's identity that, for all $0\leq s<u<t\leq T$,
\begin{equation}\label{multiplicativesignatN}
S(X)^{\leq N}_{s,t} = S(X)^{\leq N}_{s,u} \otimes S(X)^{\leq N}_{u,t}.
\end{equation}
We say that the truncated signature $S(X)^{\leq N}$ is \emph{multiplicative}. 

\subsection{Rough Paths} 

A rough path can be viewed as a \emph{lift} of a path, obtained by enriching it with its iterated integrals. For smooth paths, this coincides with the classical signature, while for irregular paths this structure needs to be defined abstractly. We start with the following definition, that extends (\ref{multiplicativesignatN}) to a more general setting. 

\begin{defi}[Multiplicative functional]
    For $N \in \N$, let $\bX : \Delta_T \to T^N(V)$ be a continuous map and denote $\bX(s,t)=\bX_{s,t}$. Since $\bX_{s,t} \in T^N(V)$, we can write $\bX_{s,t} = (\bX_{s,t}^0, \bX_{s,t}^1, \dots, \bX_{s,t}^N)$, where $\bX_{s,t}^k \in V^{\otimes k}$ for each $k$. We say that $\bX$ is a \textnormal{multiplicative functional} of degree $N$ in $V$ if, for every $(s,t)\in \Delta_T$, we have $\bX_{s,t}^0:=1$ and 
    \begin{equation} \label{chenrough}
    \bX_{s,t} = \bX_{s,u} \otimes \bX_{u,t}
    \end{equation}
    for all $s\leq u\leq t$.
\end{defi}

By extension, we also refer to \eqref{chenrough} as Chen's identity. Instead of constructing the signature from a classical path, we now assume the algebraic structure of a signature and study its properties as a path taking values in the tensor algebra $T^N(V)$. 

For a continuous functional $\bX:\Delta_T\to T^N(V)$, let 
\begin{equation}\label{p_var_norm}
    ||\bX||_{p\cdot \text{var}}:=\max_{1\leq k\leq N} \sup_{\mathcal{D}}{\left( \sum_{t_i\in\mathcal{D}} ||\pi_k(\bX_{t_i,t_{i+1}})||_{V^{\otimes k}}^{p/k} \right)^{k/p}},
\end{equation}
where the $\sup$ is taken over all the partitions $\mathcal{D}_{[0,T]}$. If $||\bX||_{p\cdot \text{var}}<\infty$, the functional $\bX$ is said to have finite $p$-variation. The $p$-variation distance between two functionals $\bX$ and $\bY$ of  finite $p$-variation is defined as 
\begin{equation}\label{pvardistance}
    d_{p\cdot \text{var}}(\bX,\bY):=||\bX-\bY||_{p\cdot \text{var}}.
\end{equation}

The following result highlights that, for a multiplicative functional with finite $p$-variation, the first $\floor{p}$ levels completely capture all of its information.

\begin{teo}[Lyons' Extension Theorem]\label{extensionthm} 
Let $p \geq 1$ be a real number, $N \geq 1$ an integer, and let $\bX : \Delta_T \to T^N(V)$ be a multiplicative functional of degree $N$ with finite $p$-variation. Suppose that $N \geq \floor{p}$. Then, for every integer $n > N$, there exists a unique continuous multiplicative functional 
\[
\bY : \Delta_T \to T^n(V) 
\]
such that $\bY$ agrees with $\bX$ up to level $N$ (that is, $\pi_{\leq N}(\bY) = \bX$), and $\bY$ has finite $p$-variation. Moreover, the map that sends $\bX$ to its extension $\bY$ is continuous with respect to the $p$-variation metric.
\end{teo}

This important result allows us to work with “compressed” versions of multiplicative functionals, which contain exactly the levels up to $\floor{p}$ because they capture all relevant information. 

\begin{defi}[Rough Path]
    Let $p \geq 1$. A \emph{$p$-rough path} is a continuous multiplicative functional 
    \[
    \bX : \Delta_T \to T^{\floor{p}}(V)
    \]
    of degree $\floor{p}$ with finite $p$-variation. The space of $p$-rough paths is denoted by $\Omega^p_T(V)$.
\end{defi}

Rough paths form a highly abstract class, while paths of bounded variation are concrete and familiar. It is therefore natural to examine the rough paths that are \emph{close} to signatures of bounded variation paths, namely, those that arise as limits of such signatures. 

\begin{defi}[Geometric Rough Paths] A \emph{geometric $p$-rough path} is a $p$-rough path $\bX$ for which there exists a sequence of paths of bounded variation $(X_n)_{n \geq 1}$ such that
\[
\lim_{n \to \infty} d_{p\cdot \textnormal{var}}(\bX, S(X_n)^{\leq \floor{p}}) = 0.
\]
The space of geometric $p$-rough paths is denoted by $G\Omega^p_T(V)$.
\end{defi}

It follows from Proposition \ref{grouplikeprop} that the signature of a bounded variation path satisfies the group-like property. Therefore, every geometric $p$-rough path takes values in $G^{\floor{p}}(V)$, the set of group-like elements in the truncated tensor algebra $T^{\floor{p}}(V)$.

However, not every $p$-rough path taking values in $G^{\floor{p}}(V)$ arises as the limit of signatures of bounded variation paths. This distinction motivates the following definition.

\begin{defi}[Weakly Geometric Rough Paths]
    A \emph{weakly geometric $p$-rough path} is a $p$-rough path taking values in $G^{\floor{p}}(V)$. The space of weakly geometric $p$-rough paths is denoted by $WG\Omega^p_T(V)$.
\end{defi}

Since the Itô signature of Brownian motion fails to be \emph{group-like}, the canonical rough path lift is taken to be the \emph{Stratonovich lift}, which is group-like: 
\[
\mathbf{B}_{s,t} := \left(1,\, B_t - B_s,\, \int_s^t (B_u - B_s) \circ dB_u \right).
\]

\subsection{Time-augmented Paths}

Let $S(X)$ denote the signature of a path $X : [0,T] \to V$. It is well known that the map $X \mapsto S(X)$ is not injective: different paths can share the same signature—for example, if they trace out the same image at different speeds. To address this issue, instead of lifting $X_t$ alone, we lift the augmented path $\hat{X} : [0,T] \to \R \oplus V$ defined by:
\[
\hat{X}_t:= (t, X_t).
\] 
Proposition~\ref{augmentXtohatX} below shows that, under time augmentation, a path that admits a lift to a weakly geometric $p$-rough path continues to do so. And Theorem~\ref{signatuniqueness} shows that the time-augmented signature uniquely determines the path (i.e., that augmenting the path by adding time recovers injectivity).
\begin{prop}\label{augmentXtohatX}
Let $X: [0,T] \to V$ be a continuous path that admits a weakly geometric $p$-rough path lift $\bX \in WG\Omega^p_T(V)$. Define the time-augmented path $\hat{X}:[0,T] \to \R \oplus V$ by
\[
\hat{X}_t := (t, X_t).
\]
Then $\hat{X} $ also admits a weakly geometric $p$-rough path lift $\hat{\bX} \in WG\Omega^p_T(\R \oplus V)$.
\end{prop}
Once we know that there exists a lift for the time-augmented path $\hat{X}$, we define the corresponding space. 

\begin{defi}[Time-Augmented Weakly Geometric Rough Path]\label{DEFaugmentedWGpRP}
A \emph{time-augmented weakly geometric $p$-rough path} is a weakly geometric $p$-rough path $\bX \in WG\Omega^p_T(\R \oplus V)$ such that
\begin{itemize}
    \item [(i)] the first level satisfies $\pi_1(\bX_{s,t}) = \hat{X}_t - \hat{X}_s$, where $\hat{X}_t = (t, X_t)$ for some continuous path $X: [0,T] \to V$ that admits a weakly geometric $p$-rough path lift;
    \item [(ii)] for any $I$ with $|I|\leq \lfloor p\rfloor$,
                \[
                \langle e_{I0},\bX_{s,t} \rangle = \int_s^t \langle e_I, \bX_{s,u} \rangle du,
                \] 
    where the integral is a Young integral, and where the notation $e_{I0}$ means appending a $0$ (the time component) to the multi-index $I$.
\end{itemize}
We denote the space of such paths by $WG\hat{\Omega}^p_T(V)$.
\end{defi}

Time augmentation allows the signature to distinguish between paths that follow the same geometric shape but evolve at different speeds; it is essential for learning path-dependent functionals in an \emph{adapted} way. Without time, the signature treats two paths with the same shape but different timing as indistinguishable—an undesirable feature in many stochastic or temporally sensitive learning tasks.

Note that a rough path in $WG\hat{\Omega}^p_T(V)$ presupposes the existence of a path $X:[0,T] \to V$, which is augmented to $\hat{X}:[0,T] \to \R \oplus V$. This contrasts with the rough paths in $WG\Omega^p_T(V)$, which are defined independent of any base path. We now proceed to extend the rough paths.

\subsection{Signatures of Rough Paths}

Let $\bX \in \Omega^p_T(V)$ be a $p$-rough path. By the extension theorem, for every integer $N \geq \floor{p}$, there exists a unique multiplicative extension of $\bX$ to degree $N$ with finite $p$-variation. Since this extension process can be carried out to arbitrarily high degrees, it is natural to define the \emph{signature} of a $p$-rough path as its formal infinite extension. This motivates the following definition.

\begin{defi}[Signature of a Rough Path]\label{signatRoughPath}
Let $\bX \in \Omega^p_T(V)$ be a $p$-rough path. The \emph{truncated signature of order $N \geq \floor{p}$} is defined as the unique extension of $\bX$ to level $N$ with finite $p$-variation, denoted by:
    \[
    S(\bX)^{\leq N} := \left(1, S(\bX)^1, \dots, S(\bX)^N \right) \in T^N(V).
    \]
    The \emph{(full) signature} of $\bX$ is the formal series
    \begin{align*}
    S(\bX): \Delta_T & \to T((V)) \\
    (s,t) & \mapsto S(\bX)_{s,t} := (1, S(\bX)^1_{s,t},\dots,S(\bX)^n_{s,t},\dots),
    \end{align*}
    where $S(\bX)^n_{s,t} \in V^{\otimes n}$ denotes the $n$-th level of the extension.
\end{defi}

\noindent Note the slight shift in notation that has taken place here. We started with a path $X : [0,T] \to V$ and constructed its signature, denoted by $S(X) : \Delta_T \to T((V))$. In contrast, we now start with a multiplicative functional $\mathbf{X} : \Delta_T \to T^{\lfloor p \rfloor}(V)$, which satisfies specific algebraic and analytic properties (multiplicativity and finite $p$-variation) that allow it to be uniquely extended to higher levels $N \geq \lfloor p \rfloor$. The above definition simply relabels this extended functional using the signature notation previously introduced for bounded variation paths.

\medskip

Although we do not perform model calibration in this paper, we briefly present the universal approximation theorems to illustrate the power of signatures for representing and learning functionals of paths, particularly in supervised learning settings.

The following result states that the signature of a time-augmented weakly geometric $p$-rough path $\bX$, evaluated at time $T$, uniquely determines $\bX$. We use the simplified notation $\bX_t:= \bX_{0,t}$.

\begin{teo}[Uniqueness of the Signature] \label{signatuniqueness}
    Let $\bX$, $\bY \in WG\hat{\Omega}^p_T(V)$. Then,  
    \[
    S(\bX)_T = S(\bY)_T \iff \forall t \in [0,T],\ \bX_t = \bY_t. 
    \]
\end{teo}

Note that $WG\hat{\Omega}^p_T(V)$, equipped with the $p$-variation distance, becomes a topological space whose topology is induced by the $p$-variation metric. The following important result is proved by applying the Stone–Weierstrass theorem.

\begin{teo}[First Universal Approximation Theorem] \label{1UAT}
Let $K \subset WG\hat{\Omega}^p_T(V)$ be compact, and let $f:WG\hat{\Omega}^p_T(V)\to \R$ be continuous with respect to the $p$-variation topology. Then, for every $\eps > 0$, there exists $\ell \in T(\R\oplus V)$ such that
\[
\sup_{\bX \in K} |f(\bX) - \langle \ell, S(\bX)_T \rangle | < \eps.
\]
\end{teo}

The signature $S(\bX)$, evaluated at time $T$, serves as a feature map that transforms a path into an infinite sequence of coordinates capturing all relevant information. Theorem~\ref{signatuniqueness} ensures that this representation is injective for time-augmented paths. Theorem~\ref{1UAT} (First UAT) states that continuous functionals on compact subsets of the rough path space can be approximated arbitrarily well by linear functionals on the signature, that is, by finite linear combinations of iterated integrals. 

Consider a stochastic process of the form
\[
Y_t = f\big((\bX_s)_{s \in [0,t]}\big),
\]
with each $\bX_s \in WG\hat{\Omega}^p_t(V)$. Assume that $X:[0,s]\to V$ admits a lift $\bX \in WG\hat{\Omega}^p_{s}(V)$. We need this lift to be understood as living in $WG\hat{\Omega}^p_{t}(V)$, so that $f$ can act on a common space. To interpret a rough path in $WG\hat{\Omega}^p_{s}(V)$ as an element of $WG\hat{\Omega}^p_{t}(V)$ in a consistent way we need to stop the path (but not the time), and impose a specific topology in the new space.

The concept of stopped rough paths provides a useful framework for handling adaptedness, but constructing the corresponding space of weakly geometric \emph{stopped} $p$-rough paths, together with a suitable topology, is rather technical. We refer the reader to \cite{kalsi20} and \cite{bayer23} for a detailed treatment.  

At an intuitive level, the Second Universal Approximation Theorem extends Theorem~\ref{1UAT} to functionals depending on the path up to variable times. In particular, it ensures that any continuous functional of a stopped rough path can be uniformly approximated—over both paths and truncation times $t \in [0,T]$—by linear functionals of the corresponding truncated signatures.

We now turn to the practical implementation of these ideas. In particular, we focus on computing the signatures of volatility paths, which will serve as feature representations in the learning framework developed below.

\section{Volatility Models and their Signatures} \label{sec:vol_models}

We begin by simulating sample paths from three stochastic processes commonly used to model volatility. The Heston model is a classical stochastic volatility model in which the variance follows a mean-reverting square-root process:
\begin{equation} \label{model:Heston}
dS_t = S_t \sqrt{v_t}\, dB_t \qquad \quad dv_t = \kappa(\theta - v_t)\,dt + \nu \sqrt{v_t}\, dW_t,
\end{equation}
where $B_t$ and $W_t$ are correlated Brownian motions, and the risk-free rate is assumed to be zero. The parameters \(\kappa\), \(\theta\), and \(\xi\) respectively control the speed of mean reversion, the long-term variance level, and the volatility of volatility. The Heston model remains one of the most widely used frameworks in quantitative finance due to its analytical tractability and its ability to capture key facts such as volatility clustering.

Any continuous semimartingale $S = A + M$, where $M$ is a continuous local martingale and $A$ is a continuous path of bounded variation on compact intervals, admits a canonical Stratonovich rough path lift, which is weakly geometric (see \cite{frizvictoir10}, Chapter~14).\footnote{The roughness of $S$ is driven by the martingale component $M$, while the bounded variation part $A$ can be handled using classical integration. The construction of the lift relies on the Burkholder--Davis--Gundy inequality and properties of quadratic variation.} Therefore, the Heston variance process admits a canonical weakly geometric $p$-rough path lift. Note that the Feller condition $2\kappa \theta \geq \nu^2$ ensures positivity, but is not required for existence of the lift.

Rough volatility models, such as those in \cite{ALV}, \cite{fukasawa17}, \cite{GatheralJaiRosen18}, have gained popularity because they capture key empirical features of realized volatility that traditional Markovian models struggle to reproduce (such as the exploding term structure of the at-the-money skew as maturity tends to zero) and, when embedded in pricing models, produce option prices that match market data remarkably well. Here we consider the rough Bergomi model, a non-Markovian model introduced in~\cite{bayerfrizgatheral16} and designed to reproduce the empirically observed roughness of volatility. The variance process is given by
\begin{equation}  \label{model:rBergomi}
v_t = v_0 \exp\left( \eta W_t^{H} - \tfrac{1}{2} \eta^2 t^{2H} \right),
\end{equation}
where \(W_t^{H}\) is a fractional Brownian motion with Hurst parameter $H < 1/2$. 

This model captures the fact that log-volatility exhibits rough sample paths, with regularity lower than that of Brownian motion. Empirical studies have shown that such rough volatility models provide a better fit to market data, particularly in reproducing the short-maturity implied volatility. See, for example, \cite{GatheralJaiRosen18}, \cite{livieri18} and \cite{bayer17}.

Unlike the case of continuous semimartingales, the rough path lift of fractional Brownian motion must be constructed using techniques from Gaussian process theory. See~\cite{FrizHairer24}, Chapter 10. \cite{CoutinQian02} show that fractional Brownian motion admits a canonical geometric rough path lift for any Hurst parameter $H > 1/4$.

Finally, we also consider the Ornstein--Uhlenbeck (OU) process, defined by
\begin{equation}  \label{model:OU}
dX_t = \kappa(\theta - X_t)\,dt + \sigma\, dW_t,
\end{equation}
which is a mean-reverting Gaussian process. While the OU process is not a realistic model for volatility in modern financial markets—since it lacks both stochastic volatility structure and roughness—it provides a useful benchmark due to its simplicity and well-understood properties.

As the Ornstein--Uhlenbeck process is a continuous semimartingale, it admits a canonical weakly geometric rough path lift in the Stratonovich sense (it does not exhibit any additional roughness beyond that induced by the Brownian motion).

We primarily focus on the Heston and rough Bergomi models, as they represent the most relevant frameworks for capturing market volatility dynamics in modern quantitative finance. The Ornstein--Uhlenbeck process is only included to test the ability of our methodology to distinguish between different types of dynamics. 

If $V$ has dimension $d$, let $d_N := \sum_{k=0}^N d^k$ denote the dimension of the truncated tensor algebra $T^N(V)$. The following notation allows us to represent elements of $T^N(V)$ as vectors in $\R^{d_N}$. Let
\[
\mathcal{L} : \{ I;\ |I| \leq N \}\ \to\ \{1,\dots,d_N\}
\]
be a \emph{labeling} function, that is, a bijection that assigns a unique index to each multi-index $I$ of length at most $N$. For any $\ell = \sum_{|I| \leq N} \ell_I e_I$, the map
\[
\begin{array}{rccl}
\mathbf{vec} : & T^N(V) & \to & \R^{d_N} \\[1ex]
& \ell & \mapsto & \big( \ell_{\mathcal{L}^{-1}(1)}, \dots, \ell_{\mathcal{L}^{-1}(d_N)} \big)
\end{array}
\]
\emph{flattens} the elements of $T^N(V)$, making it possible to identify tensors with vectors in $\R^{d_N}$. This makes the use of signature data in numerical algorithms quite convenient. 

For example, assume that the dynamics of $(X_t)_{t\geq 0}$ are given by 
\begin{equation} \label{XprocessExample}
    dX_t = \kappa(\theta - X_t)\, dt + \nu \sqrt{X_t}\, dW_t.
\end{equation}
By Proposition~\ref{augmentXtohatX}, the time-augmented process also admits a weakly geometric $p$-rough path lift $\bX \in WG\Omega^p_T(\R \oplus \R)$, where time corresponds to coordinate $0$ and $X_t$ to coordinate $1$. The truncated signature of order~$2$ has the form
\[
S(\bX)_t^{\leq 2} = \left(1,\; (t, X_t),\;
\begin{pmatrix}
\int_0^t s\, ds & \int_0^t s\, dX_s \\
\int_0^t X_s\, ds & \int_0^t X_s\, dX_s
\end{pmatrix}
\right),
\]
and the \emph{flattened} signature can be written as:
\[
\mathbf{vec}(S(\bX)^{\leq 2}_t) = \left(1,\, t,\, X_t,\, \frac{1}{2}t^2,\, \int_0^t s\,dX_s,\, \int_0^t X_s\,ds,\, \int_0^t X_s\, dX_s  \right)
\]
The third level of the signature, $S(\bX)_t^3$, consists of iterated integrals over $0 < u_0 < u_1 < u_2 < t$, indexed by $I = (i_0, i_1, i_2) \in \{0,1\}^3$, and it includes $2^3 = 8$ terms:
\[
S(\bX)_t^3\ =\ 
\left(
    \begin{array}{ll}
        \int_0^t \int_0^{u_2} \int_0^{u_1} du_0\, du_1\, du_2
        & \int_0^t \int_0^{u_2} \int_0^{u_1} du_0\, du_1\, dX_{u_2} \\[1ex]
        \int_0^t \int_0^{u_2} \int_0^{u_1} du_0\, dX_{u_1}\, du_2
        & \int_0^t \int_0^{u_2} \int_0^{u_1} du_0\, dX_{u_1}\, dX_{u_2} \\[1ex]
        \int_0^t \int_0^{u_2} \int_0^{u_1} dX_{u_0}\, du_1\, du_2
        & \int_0^t \int_0^{u_2} \int_0^{u_1} dX_{u_0}\, du_1\, dX_{u_2} \\[1ex]
        \int_0^t \int_0^{u_2} \int_0^{u_1} dX_{u_0}\, dX_{u_1}\, du_2
        & \int_0^t \int_0^{u_2} \int_0^{u_1} dX_{u_0}\, dX_{u_1}\, dX_{u_2}
    \end{array}
\right).
\]
The flattened truncated signature of order~$3$ is thus a vector in $\mathbb{R}^{15}$. In Sections~\ref{sec:numerical_fixed} and~\ref{sec:numerical_random}, we work with truncated signatures of order~$4$, for which $d_N = 31$, so that the vectorized signature $\mathbf{vec}(S(\mathbf{X})^{\leq 4}_t)$ lies in $\mathbb{R}^{31}$. Although this corresponds to a higher-order feature representation, the resulting dimension remains modest: each path is summarized by only $31$ features, rather than by the full collection of observations along the trajectory.

\section{Gradient boosting models} \label{sec:xgboost}

Our data has a tabular structure consisting of feature vectors obtained from truncated signatures. For this type of framework, tree-based ensemble methods are known to be accurate, robust, and easy to tune. This section briefly reviews the main ideas behind gradient boosting and the XGBoost algorithm.

A \emph{decision tree} is a simple and intuitive predictive rule that makes decisions by recursively partitioning the input space. At each internal node, the data are split according to a rule of the form $x_j \leq c$, where $x_j$ is one of the input features and $c$ is a threshold. This process continues until reaching the terminal nodes (leaves), where the model outputs a prediction. Intuitively, a decision tree can be seen as a sequence of \emph{if--then} rules that partition the feature space into regions. 

To improve predictive performance, it is common to combine multiple trees (that learn weakly) into a single, more powerful model. Such approaches are known as \emph{ensemble methods}. The basic idea is that while individual trees may be unstable, their combination can yield a more accurate predictor. 

As an illustration, consider a binary classification problem where the goal is to determine whether a path comes from volatility Model~A or~B, based on two signature features $x_1$ and $x_2$. A simple decision tree might proceed as follows:
\begin{itemize}
    \item If $x_1 \leq 0.5$, predict Model~A;
    \item otherwise, check $x_2$:
    \begin{itemize}
        \item if $x_2 \leq 1.2$, predict Model~B;
        \item else, predict Model~A.
    \end{itemize}
\end{itemize}
Such a tree partitions the feature space into regions. An ensemble model improves upon this by combining many such trees, each capturing different aspects of the data. In \emph{gradient boosting}, trees are added sequentially, each correcting the errors of the current model. This stands in contrast to \emph{random forests}, another ensemble method where trees are built independently in parallel and their predictions are averaged. 

XGBoost (Extreme Gradient Boosting), introduced by \cite{chen16xgboost}, is a highly efficient and widely used implementation of gradient boosting. It incorporates several important enhancements to prevent overfitting and to allow fast training on large datasets. As with all supervised classification methods, training proceeds by minimizing a loss function that measures the discrepancy between predicted and true class labels. In the multi-class setting considered here, this is the categorical cross-entropy loss, which encourages the classifier to assign high probability to the correct model class.\footnote{The categorical cross-entropy loss is defined as $\ell(y) = -\sum_{k=1}^{K} \mathbf{1}_{\{y = k\}} \log \hat{p}_k$, where $y \in \{1, \dots, K\}$ is the true class label and $\hat{p}_k$ is the predicted probability of class~$k$.}

In our setting, we adopt XGBoost as the primary classification algorithm. Its task is to assign each volatility path to one of 
several candidate volatility model classes based on its truncated signature representation.

While the empirical machine learning literature documents a tendency for tree-based methods to perform well on tabular data 
\citep{chen16xgboost, grinsztajn22, shwartz-ziv22, WuLiMa21, giannakas21}, such comparisons should be interpreted with care. Any benchmark is necessarily limited to a specific collection of datasets, and the \emph{no-free-lunch} principle implies that no single model dominates uniformly across all problems \citep{shwartz-ziv22}. 

The strong performance of gradient boosting in our experiments is not simply a consequence of a general superiority over neural networks on tabular data, but reflects the fact that path signatures already provide a structured, geometrically meaningful representation of the input data: each component of the truncated signature corresponds to a specific iterated integral encoding a well-defined aspect of path geometry. This feature space aligns naturally with the strengths of tree-based methods, which excel at identifying nonlinear combinations of features rather than learning representations from scratch. In this sense, the signature does much of the work that a neural network would need to learn implicitly from raw path data.

\subsection{The algorithm}

We begin by simulating volatility paths from the Heston model, the rough Bergomi model for different values of $H$, and the Ornstein--Uhlenbeck model. For each simulated trajectory, we compute its truncated signature, which serves as a feature representation of the path. This corresponds to steps~$1$ and $2$ in Algorithm~\ref{algo:XGBoost} below.

The signature features are then used as input to an XGBoost classifier, which is trained to assign each path to the stochastic process from which it was generated (steps~$3$--$5$). Finally, in step~$6$ we evaluate the performance of the trained model on out-of-sample data, that is, on trajectories not used during training.

\begin{algorithm}[!htbp] 
{\small
\DontPrintSemicolon
\SetAlgoLined
\LinesNumbered
\vspace{1em}
\caption{Signature-based Model Classification with XGBoost}
\label{algo:XGBoost}

\hrule \vspace{0.3em}
\SetKwInOut{Input}{\makebox[1.2cm][r]{Input}}
\SetKwInOut{Output}{\makebox[1.2cm][r]{Output}}
\Input{Model classes $\{\mathcal{M}_k\}_{k=1}^K$; number of paths per class $n_{\mathrm{MC}}$; truncation level $N$}
\Output{Classifier $\hat{g} : \mathbb{R}^d \to \{1,\dots,K\}$ and performance metrics}

\BlankLine
\hrule
\BlankLine

\textbf{Step 1 (Data generation).}\\
For each class $\mathcal{M}_k$, simulate paths $\{X^{(k,i)}\}_{i=1}^{n_{\mathrm{MC}}}$.

\BlankLine
\hrule
\BlankLine

\textbf{Step 2 (Feature extraction).}\\
For each path $X^{(k,i)}$, compute the truncated signature
\[
\Phi^{(k,i)} := \mathbf{vec}\Big(S(\mathbf{X}^{(k,i)})^{\leq N}_t\Big) \in \mathbb{R}^{d_N}, \text{ with } d_N = \sum_{k=0}^N 2^k.
\]
\hrule
\BlankLine

\textbf{Step 3 (Dataset construction).}\\
Form the labelled dataset
\[
\mathcal{D} := \Big\{\big(\Phi^{(k,i)},\, y^{(k,i)}=k \big):\ i=1,\dots,n_{\mathrm{MC}};\; k=1,\dots,K\,\Big\}.
\]
\hrule
\BlankLine

\textbf{Step 4 (Train-test split).}\\
Partition $\mathcal{D} = \mathcal{D}_{\mathrm{train}} \cup \mathcal{D}_{\mathrm{test}}$.

\BlankLine
\hrule
\BlankLine

\textbf{Step 5 (Learning).}\\
Train an XGBoost classifier to learn the mapping $\hat{g} : \Phi \mapsto y$ by solving
\[
    \hat{g} \;\in\; \arg\min_{g \in \mathcal{G}} 
    \sum_{(\Phi,y)\in \mathcal{D}_{\mathrm{train}}} \ell\big(g(\Phi),\, y\big),
\]
where $\ell$ is a categorical cross-entropy loss function.

\BlankLine
\hrule
\BlankLine

\textbf{Step 6 (Prediction and evaluation).}\\
Evaluate $\hat{g}$ on $\mathcal{D}_{\mathrm{test}}$ and compute the confusion matrix.

\BlankLine
\hrule \vspace{1.3em}
}   
\end{algorithm}

\vspace{0.4em}
For each model, we generate $250{,}000$ sample paths, resulting in a total dataset of one million trajectories. The data are split into training and test sets, with $n_{\mathrm{test}} = 50{,}000$ paths reserved for out-of-sample evaluation in each model. The dataset is therefore well balanced across classes, avoiding issues associated with class imbalance and ensuring that the classification performance is not biased toward any particular model.

\paragraph{Implementation details.}
All computations are carried out in Python. The simulation of rough Bergomi paths follows the hybrid scheme of \cite{BLPakkanen2017}, and builds on the code of \cite{McCrickturbo18},\footnote{Available at \url{https://github.com/ryanmccrickerd/rough\_bergomi}.} adapted for GPU acceleration. 

\noindent Signature computations are based on a vectorized adaptation of the implementation by Peter Foster,\footnote{Available at \url{https://github.com/pafoster/path\_signatures\_introduction}.} modified for GPU execution. 

\noindent The XGBoost classifier is implemented using the standard \texttt{XGBClassifier} from the \texttt{xgboost} Python package \citep{chen16xgboost}, with a learning rate of $0.05$, maximum tree depth of~$6$, and $500$ estimators; GPU acceleration is enabled to handle the large dataset efficiently. 

\noindent The remaining components of the pipeline---path generation for the Heston and OU models, dataset construction, and train-test splitting---are implemented from scratch using standard NumPy and CuPy routines. The full code will be made available upon publication.

\vspace{1em}

We do not use a separate validation set for the XGBoost experiments, since overfitting is controlled directly through built-in regularization mechanisms, which provide stable out-of-sample performance without requiring an explicit early-stopping procedure. 

Note that, although random parameter sampling introduces substantial within-class variability, the paths remain realizations of 
well-specified stochastic models, and extreme or anomalous observations arise with probabilities governed by those models. With real market data, by contrast, structural breaks, liquidity distortions, and other microstructure effects can produce outliers that are qualitatively different from anything seen during training---in other words, with synthetic data even the tails are part of the model, whereas with market data, outliers may lie outside the model's support entirely. As a result, model-generated data provides a more controlled learning environment than empirical time series, and the generalization gap between training and test performance is expected to be smaller than it would be with real market data.

To evaluate performance, we use the confusion matrix and overall classification accuracy, defined as the proportion of correctly classified paths. This is the most natural performance measure in our setting, where the goal is to identify the underlying volatility model from an observed path. Standard binary metrics such as false positive and false negative rates are less informative here, since the problem is inherently multi-class and does not involve a decision threshold in the usual sense.

\section{Classification with constant model parameters} \label{sec:numerical_fixed}

This section serves as a \emph{proof of concept} to assess whether signature features can distinguish volatility dynamics in controlled settings, where each model class is simulated using fixed parameter values. Consequently, the boosting algorithm does not classify model classes in general, but rather trajectories generated from models with a specific parameter configuration.

For the Heston variance
\[
dX_t = \kappa(\theta - X_t)\,dt + \nu \sqrt{X_t}\, dW_t,
\]
we consider $X_0 = 0.1,\, \kappa = 2,\, \theta = 0.1,\, \nu = 0.2$. For the rough Bergomi variance
\[
X_t = \xi \exp\left( \eta W_t^{H} - \tfrac{1}{2} \eta^2 t^{2H} \right), 
\]
we set $\xi = 0.08$ and $\eta = 1.8$. In each experiment we consider several values of $H$, representing different degrees of roughness. Finally, for the Ornstein--Uhlenbeck process 
\[
dX_t = \kappa(\theta - X_t)\,dt + \sigma\, dW_t,
\]
we use parameters $X_0 = 0.15,\, \kappa = 3,\, \theta = 0.15,\, \sigma = 0.1$.

Paths are simulated over a short time horizon $T = 0.1$ in order to capture roughness features, which are most pronounced at short time scales. The truncated signatures are computed from discretized paths with $100$ time steps and used as feature vectors for the learning algorithm.

In the numerical computations below, we focus on relatively small values of the Hurst parameter~$H$, in line with the empirical evidence reported in the literature. In particular, \cite{GatheralJaiRosen18} document that log-volatility exhibits rough behavior, with typical estimates of the Hurst parameter around $H \approx 0.1$. This finding is further supported by \cite{bennedsen22}, who show that volatility dynamics display distinct short- and long-term regimes, with rough behavior dominating at short time scales. \cite{WuMuzyBacry22} analyze a large set of financial time series and find that stock market indices typically have Hurst exponents in the range $H \approx 0.1$--$0.15$, while individual stocks often exhibit even smaller values. We therefore run most of our simulations for small values of $H$ in order to reflect realistic market conditions, although in this section we occasionally consider larger values as a proof of concept to explore smoother volatility dynamics.

As mentioned above, \cite{CoutinQian02} show that fractional Brownian motion admits a canonical geometric rough path lift when $H > 1/4$. Lower values of $H$, like those we use below, do not pose difficulties in practice because we work with the \emph{time-augmented} path $\hat{X}_t = (t, X_t)$, where the bounded variation time component adds regularity to the system. Besides, signatures are computed from discrete observations of the path, which are interpolated linearly, and such piecewise linear approximations always admit a canonical rough path lift.\footnote{\cite{issa23} note that the choice of interpolation method typically has little impact and that linear interpolation is often sufficient. In the context of model calibration with signatures, \cite{ABSV26} experiment with cubic splines and also report that the marginal gains in accuracy are negligible.}

\subsection{Experiment 1: Heston vs OU vs rough Bergomi} \label{exp_1}

\begin{figure}[!htbp]
\centering
\includegraphics[width=0.44\textwidth]{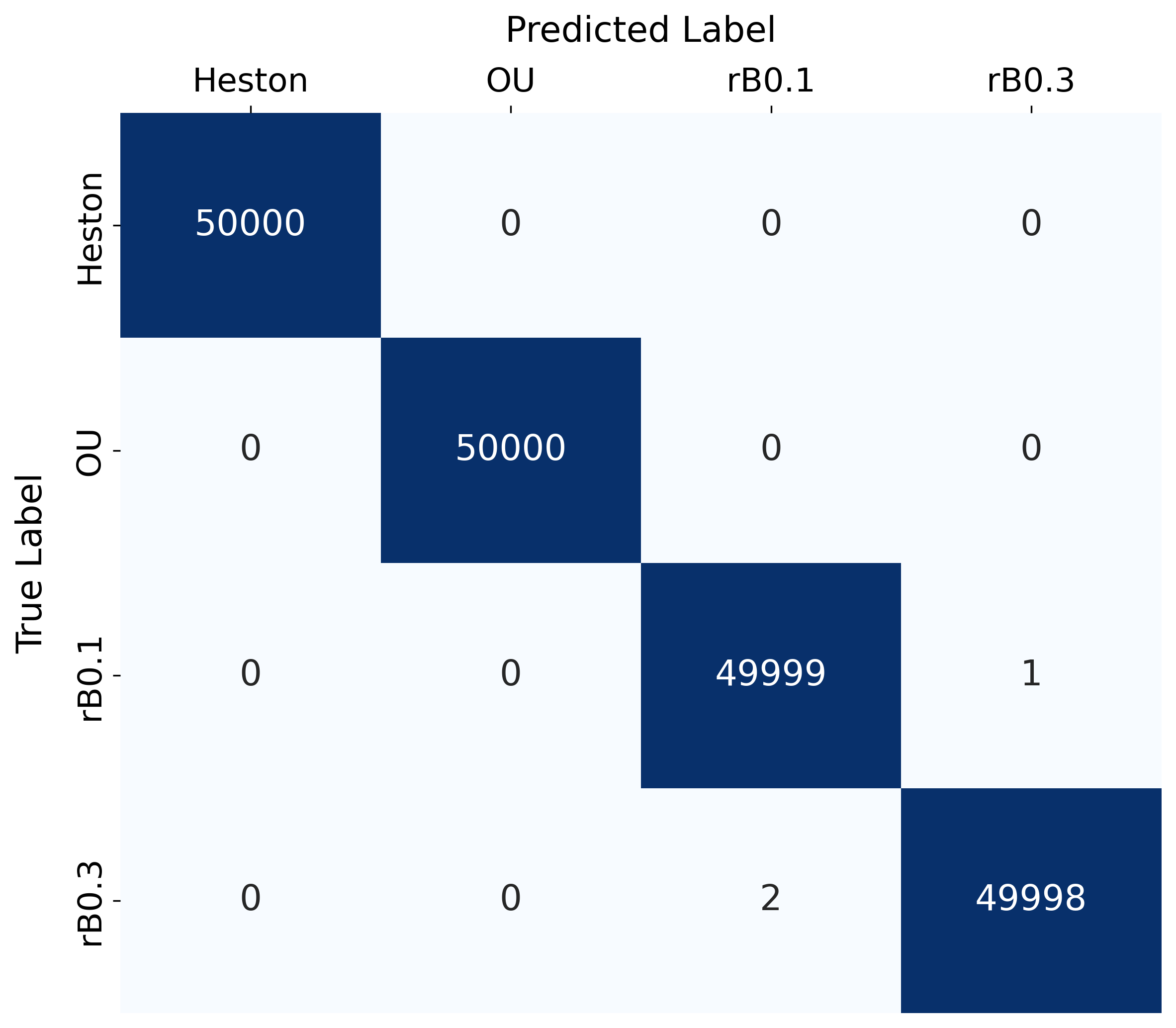}
\caption{Confusion matrix for Experiment~\thesubsection\ ($n_{\mathrm{test}} = 50{,}000$).}
\label{fig:cm_exp1}
\end{figure}

We start with a classification problem involving the Heston model, the Ornstein--Uhlenbeck process, and two rough Bergomi models with Hurst parameters $H=0.1$ and $H=0.3$. Our goal here is to check whether the learning algorithm is able to distinguish between classical and rough volatility trajectories. The classification performance is summarized by the confusion matrix in Figure~\ref{fig:cm_exp1}, where the entries correspond to the absolute number of paths in each category.

The confusion matrix indicates an almost perfect classification performance. Only three paths out of $200{,}000$ are misclassified. Interestingly, the Heston and Ornstein--Uhlenbeck processes are identified without any misclassification. 

Training and test accuracies coincide across all classes, indicating excellent generalization and no evidence of overfitting. 

In the next experiment, we extend the analysis by investigating whether the Heston volatility can be clearly distinguished from rough Bergomi models across a wider range of Hurst parameters.

\subsection{Experiment 2: Heston vs several rough Bergomi models} \label{exp_2}

\begin{figure}[!htbp]
\centering
\includegraphics[width=0.44\textwidth]{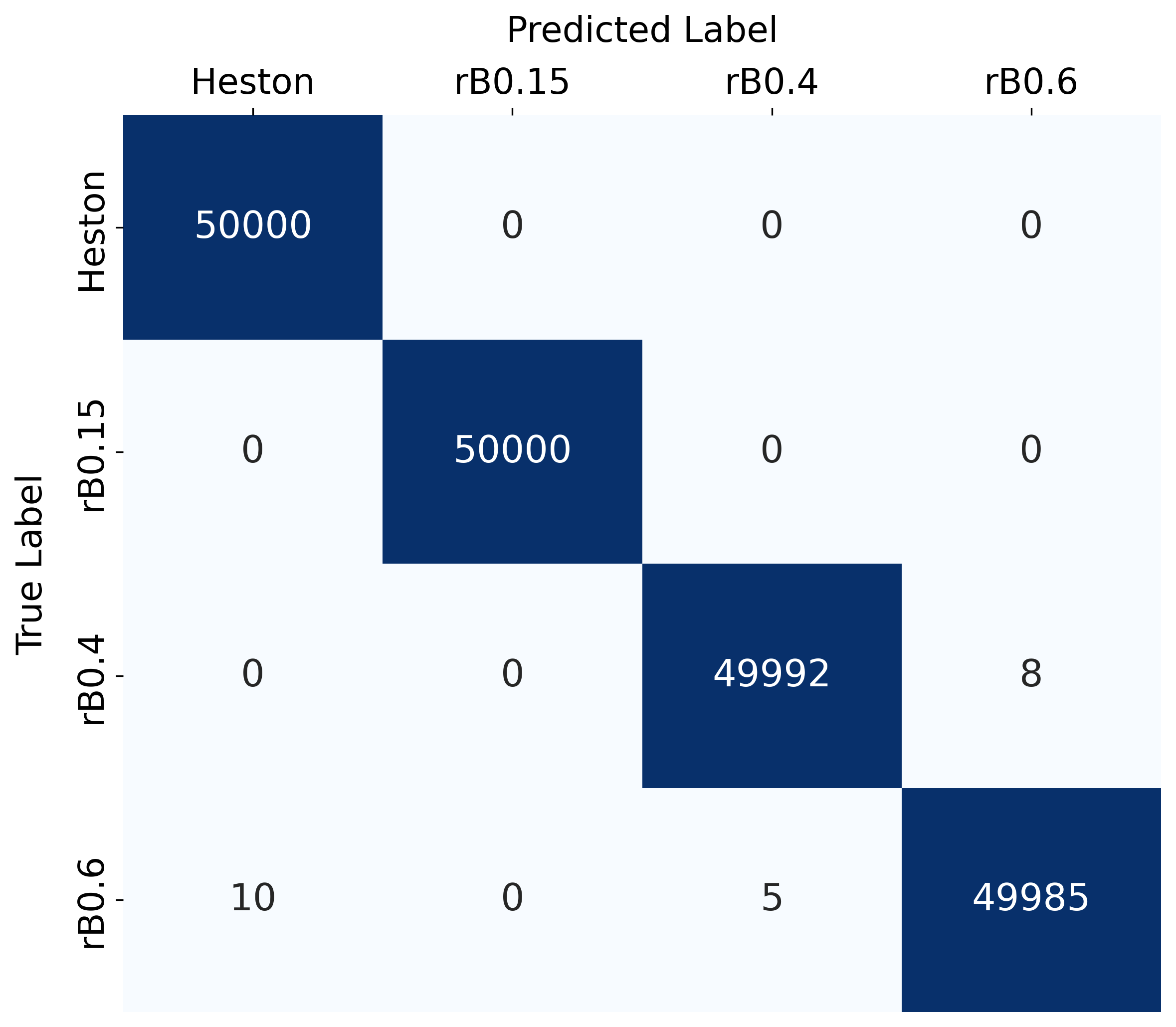}
\caption{Confusion matrix for Experiment~\thesubsection\ ($n_{\mathrm{test}} = 50{,}000$).}
\label{fig:cm_exp2}
\end{figure}

We now consider a classification problem involving the Heston and three rough Bergomi models with $H=0.15$, $H=0.4$, and $H=0.6$. In this setting, the task becomes more challenging, as the models differ more subtly in their pathwise behavior. Classification performance is summarized by the confusion matrix in Figure~\ref{fig:cm_exp2}.

The results show that the classifier is able to separate trajectories with very high accuracy, with only a negligible fraction (around $0.04\%$) of paths misclassified. Misclassifications occur between Heston and rough Bergomi trajectories with $H=0.6$, as well as between rough Bergomi trajectories with $H=0.4$ and $H=0.6$. In contrast, rough Bergomi paths with $H=0.15$ are perfectly identified, reflecting their distinctly rough behavior.

These patterns are consistent with the interpretation of the Hurst parameter: as $H$ increases, the rough Bergomi model becomes progressively smoother and closer in behavior to classical diffusion models, making the distinction more subtle. 

Importantly, the model exhibits excellent generalization properties. The training and test accuracies are nearly identical (Train: $1.0$, Test: $0.9999$), indicating that the classifier is not overfitting and generalizes effectively to unseen data. This stability is further supported by the balanced dataset and the regularization mechanisms inherent in the XGBoost algorithm.

\subsection{Experiment 3: Only rough Bergomi models} \label{exp_3}

\begin{figure}[!htbp]
\centering
\includegraphics[width=0.44\textwidth]{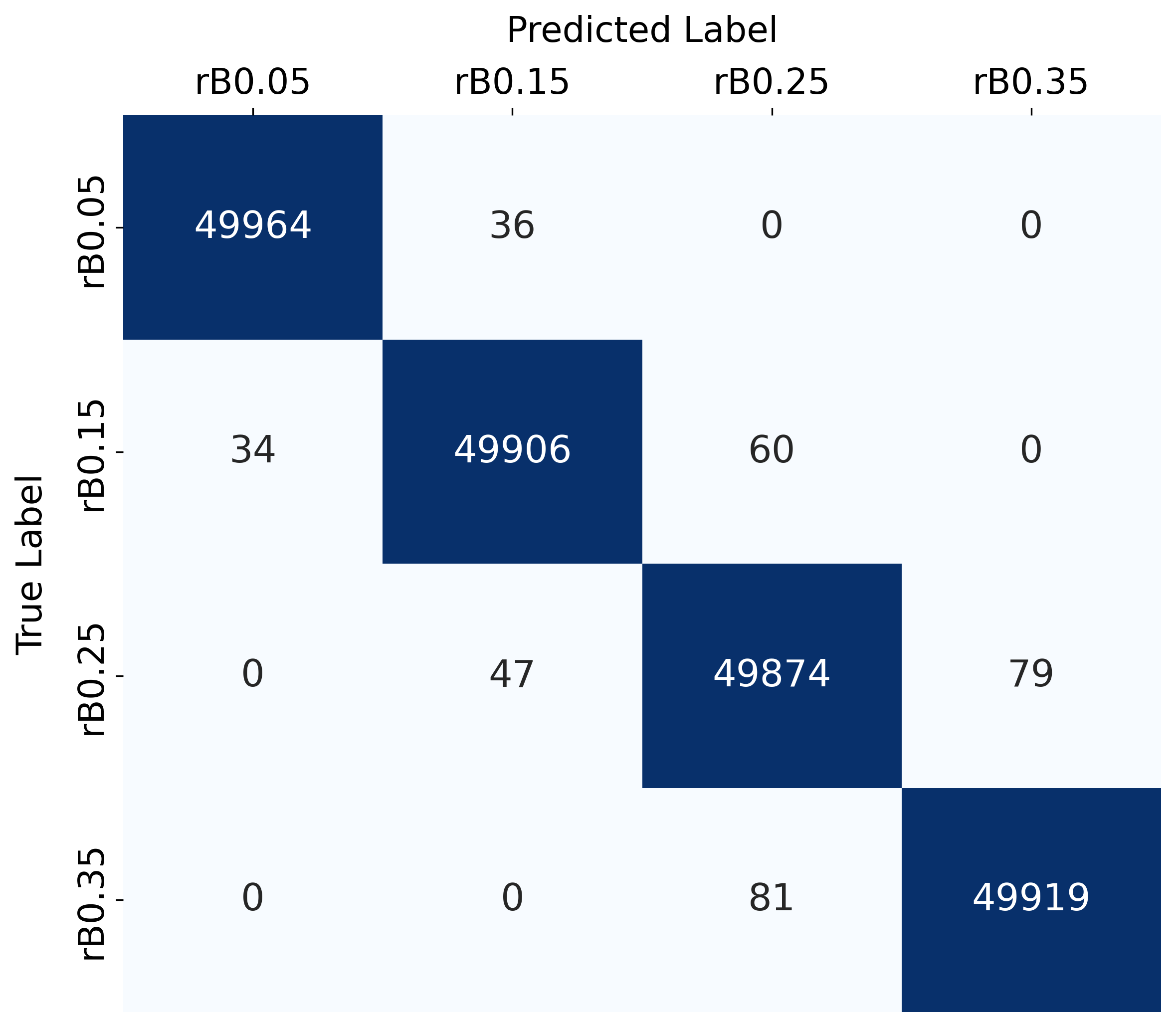}
\caption{Confusion matrix for Experiment~\thesubsection\ ($n_{\mathrm{test}} = 50{,}000$).}
\label{fig:cm_exp3}
\end{figure}

We now consider four rough Bergomi models with parameters $H=0.05$, $H=0.15$, $H=0.25$, and $H=0.35$. As all models share the same structural form and differ only through relatively close values of the Hurst parameter, we expect some misclassification, particularly between models with adjacent values of $H$. Performance is summarized in Figure~\ref{fig:cm_exp3}.

The results are as expected, but still maintaining a high classification accuracy. The confusion matrix reveals that misclassifications occur exclusively between trajectories from models with adjacent Hurst parameters: paths generated with $H=0.05$ are occasionally classified as $H=0.15$, and similarly for the pairs $(0.15,0.25)$ and $(0.25,0.35)$. In contrast, there is no confusion between trajectories from models with more distant values of $H$. And the number of misidentified paths is extremely small: the worst case ($81$ paths) represents only $0.16\%$ of the $50{,}000$ paths in its test set. This behavior is consistent with the interpretation of the Hurst parameter as a measure of path regularity. 

Importantly, the classifier continues to exhibit good generalization properties. The training and test accuracies are very close (Train: $0.9996$, Test: $0.9983$), with only a small gap between them, indicating that the model is not overfitting the training data (further confirmed by the high per-class accuracies across both training and test sets). This suggests that the XGBoost classifier, combined with signature-based features, is able to capture fine variations in path geometry while maintaining robust out-of-sample performance.

\subsection{Computational cost and practical considerations}

In Table~\ref{tab:timings} we report the running times for the different experiments. As can be seen, the dominant cost arises from the computation of the signatures, especially for the rough Bergomi paths, while the running times of the XGBoost classifier itself remain comparatively small. 

\begin{table}[H]
\centering
\begin{tabular}{lccc}
\toprule
 & Total time & Signature computation & XGBoost training \\
\midrule
Experiment~\ref{exp_1} & 0:04:09 & 0:04:01 & 0:00:08 \\
Experiment~\ref{exp_2} & 0:10:35 & 0:10:18 & 0:00:17 \\
Experiment~\ref{exp_3} & 0:15:44 & 0:15:24 & 0:00:20 \\
\bottomrule
\end{tabular}
\caption{Running times for the different experiments.}
\label{tab:timings}
\end{table}

Considering that we work with signatures truncated at order~$4$, the computational times remain moderate: the full pipeline—simulation, feature extraction, and training—can be executed within a few minutes. This is partly due to the use of vectorized implementations and GPU acceleration, which significantly reduce the computational burden. We return to the issue of computational cost in Section~\ref{sec:sample_size}.

All computations were carried out on a standard consumer desktop with $128\,\mathrm{GB}$ of RAM and an NVIDIA RTX~3080~Ti GPU. The relatively short running times highlight the practical feasibility of the approach.

\section{Classification with randomly sampled model parameters} \label{sec:numerical_random}

The experiments above were designed to assess whether signature features can distinguish volatility dynamics in controlled settings, where each model class is simulated with fixed parameter values. We now turn to a more demanding classification problem. For each simulated path, the model parameters are sampled randomly from prescribed ranges, so that the classifier must learn to identify the model class rather than a particular parameter configuration.

As in Section~\ref{sec:numerical_fixed}, paths are simulated for a maturity $T = 0.1$.

\subsection{Experiment 1: Heston vs OU vs rough Bergomi} \label{exp_rnd1}

We now repeat the first experiment from Section~\ref{sec:numerical_fixed}, but with randomly sampled parameters within each model class. For the Heston variance process
\[
dX_t = \kappa(\theta - X_t)\,dt + \nu \sqrt{X_t}\, dW_t,
\]
we fix the initial value $X_0 = 0.10$ and sample the remaining parameters independently from the uniform distributions
\[
\kappa \sim U(1.0,3.0), \qquad
\theta \sim U(0.05,0.15),
\]
while $\nu$ is sampled from a uniform distribution constrained by the Feller condition,
\[
\nu \sim U\bigl(0.15,\min(0.35,\nu_{\max})\bigr),
\]
where $\nu_{\max}=0.95\sqrt{2\kappa\theta}$ enforces the Feller condition with a $5\%$ safety margin.

For the Ornstein--Uhlenbeck process
\[
dX_t = \kappa(\theta - X_t)\,dt + \sigma\, dW_t,
\]
we fix $X_0 = 0.10$ and sample
\[
\kappa \sim U(1.5,3.5), \qquad
\theta \sim U(0.10,0.20), \qquad
\sigma \sim U(0.05,0.30).
\]

Finally, for the rough Bergomi model  
\[
X_t = \xi \exp\left( \eta W_t^{H} - \tfrac{1}{2} \eta^2 t^{2H} \right),
\]
we fix the initial value $\xi=0.10$ and sample
\[
\eta \sim U(0.8,2.0).
\]

We do not randomize the initial value of the processes because signatures are constructed from iterated integrals of increments $dX_t$, and therefore do not depend on the initial level $X_0$. Fixing the initial value removes a source of variation that is irrelevant to model identification, allowing the classifier to focus on the properties of the trajectories.

We emphasize that the same random realizations of $\eta$ are reused across the rough Bergomi classes, so that they only differ through the Hurst parameter $H$, which controls the degree of path roughness. This ensures that any successful classification is attributable solely to differences in roughness rather than to differences in the sampled parameter values. Using independent $\eta$-draws for each class would introduce an additional source of variation, making it impossible to determine whether the classifier distinguishes the models through their roughness or through incidental differences in $\eta$. We note that this design makes the identification problem strictly harder for the classifier as, conditioned on $H$, the two rough Bergomi classes are statistically identical.

In subsequent experiments, where up to four rough Bergomi classes are included simultaneously, we maintain this principle of reusing the same realizations of $\eta$ across all rough Bergomi classes.

Our guess is that XGBoost should distinguish fairly well between the Markovian and non-Markovian settings. However, we harbor some doubts about its capacity to distinguish between the Heston and OU models, because both processes are driven by mean-reverting dynamics with Brownian noise and share similar local behavior, despite the nonlinear diffusion term in the Heston variance. We therefore expect some Heston volatility paths to be classified as OU, and vice versa.

\begin{figure}[!htbp]
\centering
\includegraphics[width=0.44\textwidth]{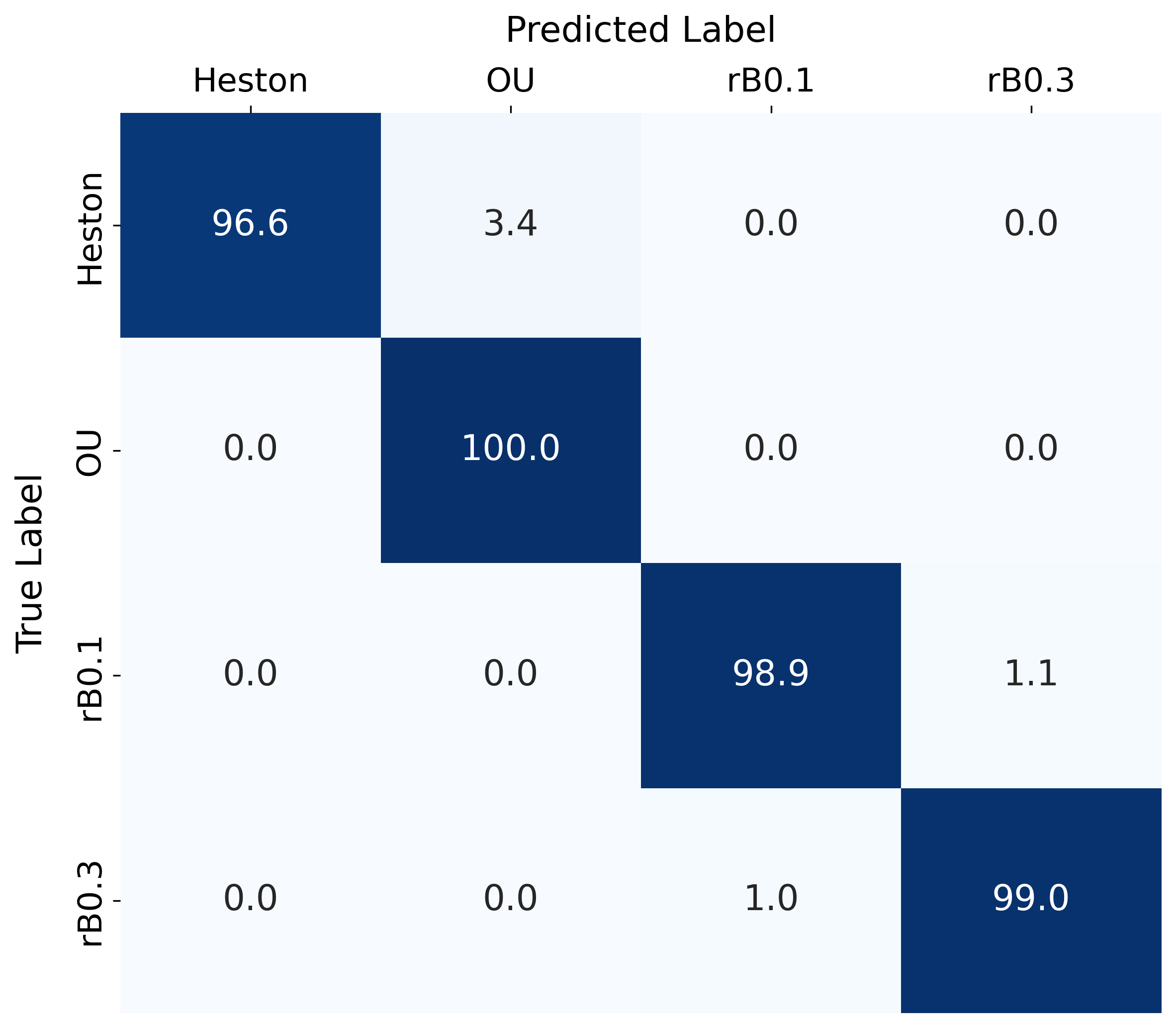}
\caption{Confusion matrix for Experiment~\thesubsection\ with randomly sampled model parameters ($n_{\mathrm{test}} = 50{,}000$). Values are percentages.}
\label{fig:cm_rnd1}
\end{figure}

The classification performance is summarized by the confusion matrix in Figure~\ref{fig:cm_rnd1}. Unlike Section~\ref{sec:numerical_fixed}, where the number of misclassified paths was extremely small and absolute counts were more informative, the confusion matrices in this section are reported in percentages in order to better highlight the classification rates across model classes.

The results show that the classifier remains highly accurate even under parameter uncertainty. The Ornstein--Uhlenbeck process is classified with 100\% accuracy, while the Heston class exhibits a misclassification rate of $3.4\%$, entirely due to confusion with the Ornstein--Uhlenbeck process. This point is examined further in a robustness check in Section~\ref{sec:OUH_robustness}.

The classifier separates the two rough Bergomi classes with accuracy close to $99\%$, providing initial evidence that signature features capture differences in path roughness in a robust manner.

\subsection{Experiment 2: Heston vs several rough Bergomi models} \label{exp_rnd2}

We now consider a more demanding classification problem involving the Heston model and three rough Bergomi classes with Hurst parameters $H=0.1$, $H=0.2$, and $H=0.3$. As the goal is to identify the underlying model class, parameters are sampled randomly within the prescribed ranges for each model.  

\begin{figure}[!htbp]
\centering
\includegraphics[width=0.44\textwidth]{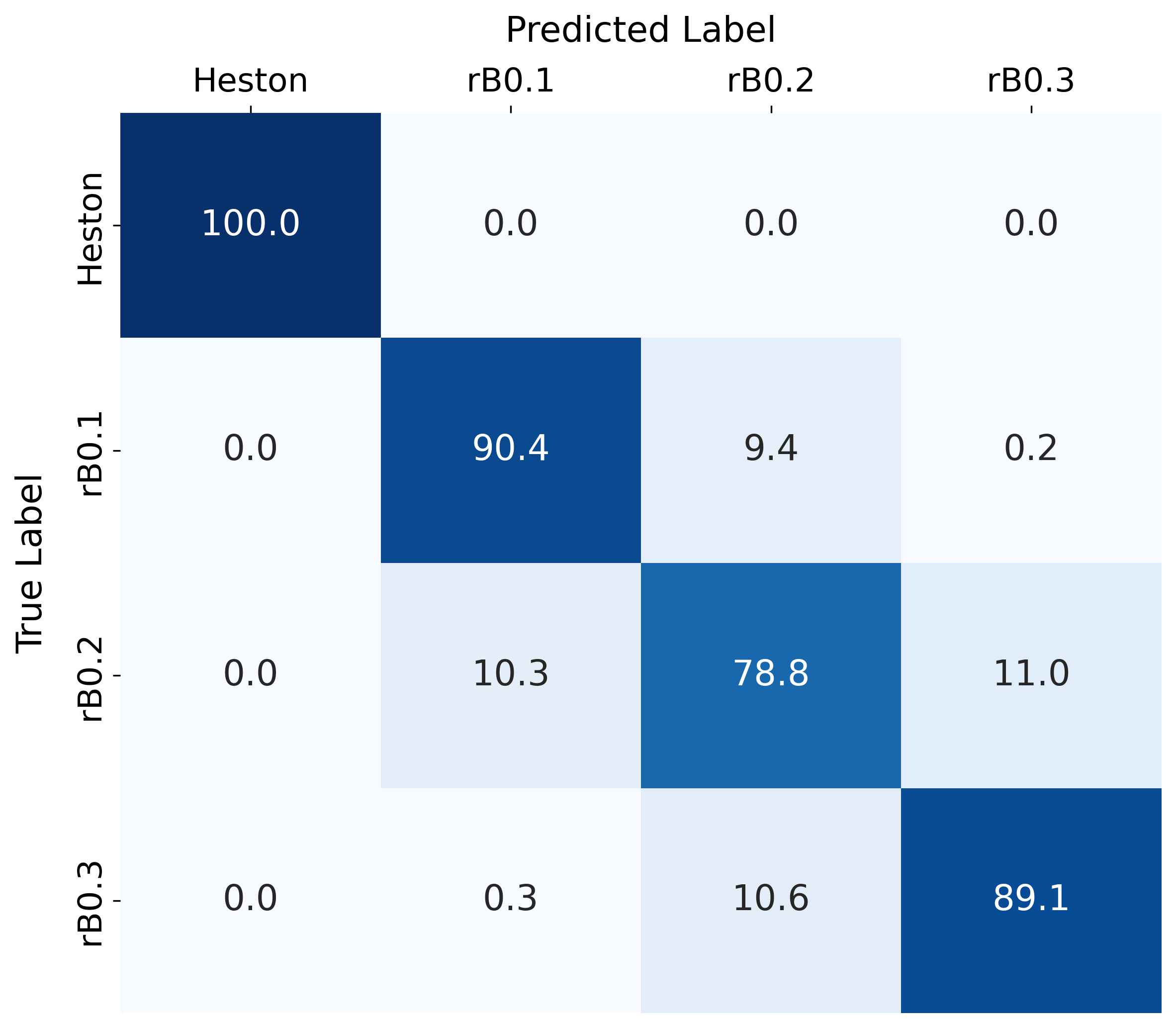}
\caption{Confusion matrix for Experiment~\thesubsection\ with randomly sampled model parameters ($n_{\mathrm{test}} = 50{,}000$). Values are percentages.}
\label{fig:cm_rnd2}
\end{figure}

The classification performance is summarized in Figure~\ref{fig:cm_rnd2}. 

The results show that the classification problem becomes substantially more challenging once parameter uncertainty is introduced and the rough Bergomi classes differ only through relatively close values of the Hurst parameter. Nevertheless, the classifier continues to exhibit strong performance. In particular, the Heston trajectories are perfectly identified, indicating that the distinction between Markovian and rough volatility dynamics remains geometrically pronounced even under random parameter sampling.

As expected, the main source of confusion occurs between the rough Bergomi classes themselves, especially between adjacent values of $H$. The model with $H=0.2$ is the most difficult to classify, as its trajectories naturally lie between the rougher regime ($H=0.1$) and the smoother regime ($H=0.3$). Misclassifications therefore occur symmetrically toward both neighboring classes. In contrast, the rough Bergomi models with $H=0.1$ and $H=0.3$ are identified with substantially higher accuracy, reflecting the more distinctive geometry of trajectories at the extremes of the roughness range considered here.

The classifier continues to exhibit good generalization properties. The gap between training and test accuracy remains small (approximately $0.5\%$), indicating that the reduction in accuracy is not caused by overfitting but rather by the intrinsic difficulty of distinguishing between closely related rough volatility classes under parameter uncertainty.

These results provide strong evidence that signature features encode information about path roughness that generalizes across parameter configurations. Even when the model parameters vary randomly from path to path, the classifier is still able to identify the underlying volatility class with high accuracy.

\subsection{Experiment 3: Only rough Bergomi models} \label{exp_rnd3}

We now consider the most challenging classification problem, involving four rough Bergomi classes with Hurst parameters $H=0.05$, $H=0.15$, $H=0.25$, and $H=0.35$. 

\begin{figure}[!htbp]
\centering
\includegraphics[width=0.44\textwidth]{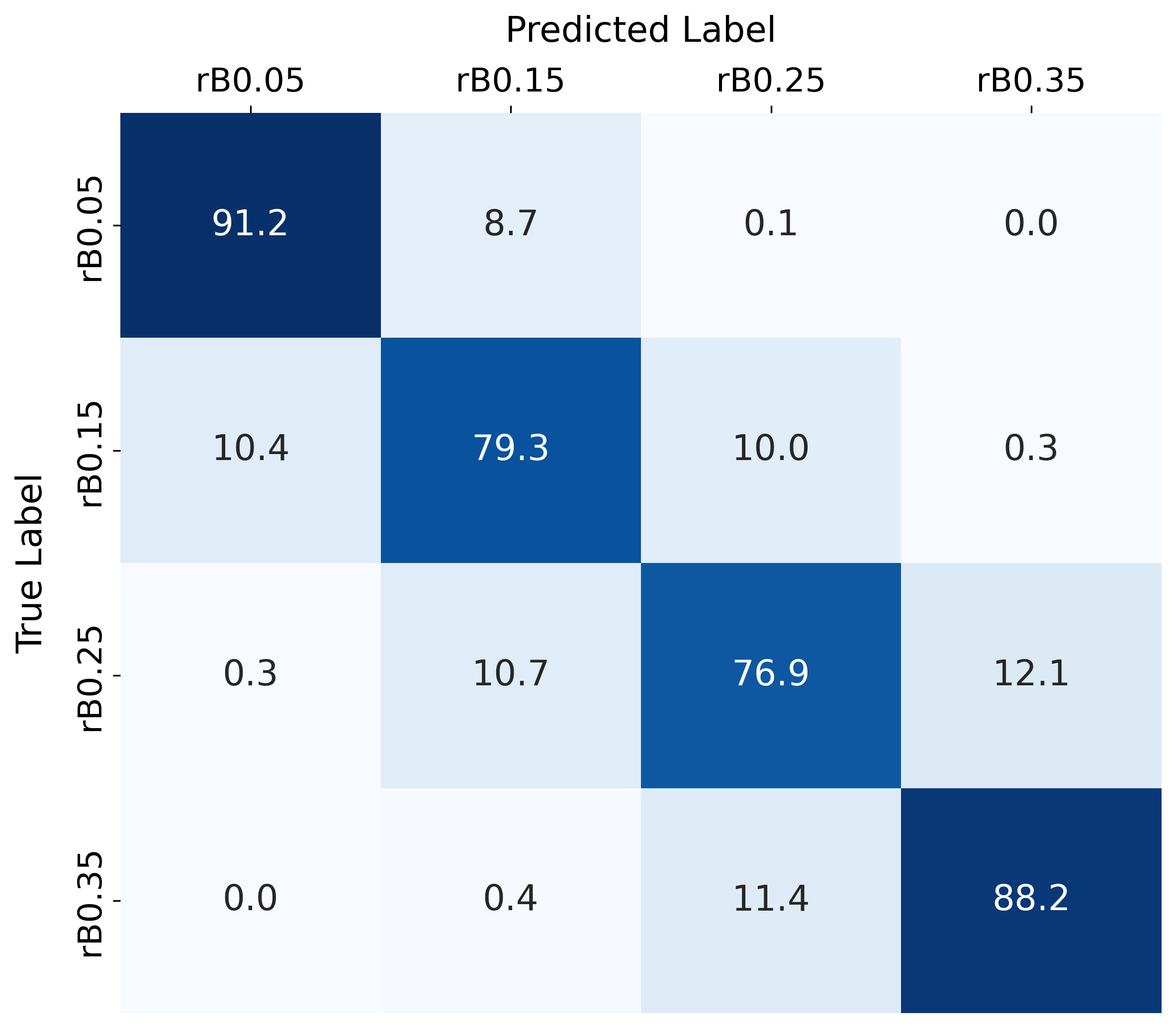}
\caption{Confusion matrix for Experiment~\thesubsection\ with randomly sampled model parameters ($n_{\mathrm{test}} = 50{,}000$). Values are percentages.}
\label{fig:cm_rnd3}
\end{figure}

Unlike Experiment~\ref{exp_rnd2}, where the classifier distinguished between Markovian and rough volatility dynamics, the present experiment focuses entirely on differences in roughness within the same model family. Successful classification relies solely on the ability of the signature features to capture increasingly subtle differences in path regularity.\footnote{Recall that across the different rough Bergomi classes, the same parameter realizations are reused for $\eta$ so that the classes differ only through the Hurst parameter.} We therefore expect the classification task to become progressively harder as the values of $H$ become closer. The corresponding confusion matrix is shown in Figure~\ref{fig:cm_rnd3}.

The results confirm that the classification problem becomes substantially more challenging when all classes belong to the same rough volatility family and the parameters are sampled randomly. In this setting, the classifier must distinguish between trajectories that differ only through relatively close values of the Hurst parameter. Nevertheless, the overall classification performance remains strong.

The confusion matrix exhibits a clear structure: misclassifications occur almost exclusively between rough Bergomi classes with adjacent values of $H$. In particular, those with $H=0.15$ and $H=0.25$ are the most difficult to distinguish, as their trajectories lie between the rougher regime ($H=0.05$) and the smoother regime ($H=0.35$). By contrast, the extreme classes are identified with higher accuracy, reflecting the more distinctive geometry of trajectories at the ends of our roughness spectrum.

The absence of significant long-range misclassifications is particularly noteworthy. For example, trajectories generated with $H=0.05$ are never classified as $H=0.35$, and vice versa. This indicates that the signature features capture a coherent notion of path roughness, where nearby values of the Hurst parameter produce geometrically similar trajectories, while more distant values remain clearly separated.

We also note that the classifier continues to generalize well despite the increased difficulty of the problem. The overall accuracy on the training and test sets is $84.7\%$ and $83.9\%$, respectively, confirming that the reduction in accuracy relative to earlier experiments is driven by the intrinsic difficulty of distinguishing closely spaced roughness levels, rather than by overfitting.

Overall, these results provide strong evidence that signature-based features are sensitive not only to broad differences between volatility models, but also to finer variations in path regularity within the same rough volatility class, even under substantial parameter uncertainty.

\subsection{Feature importance and role of signature terms} \label{sec:feature_import}

An advantage of tree-based methods is that they provide a natural measure of feature importance, quantifying how much each feature contributes to the final classification. Since each feature corresponds to one coordinate of the truncated signature, the importance scores reflect how frequently and effectively these coordinates are used in the decision trees.

After training the classifier in each experiment, we computed the built-in XGBoost feature importance scores. Across all experiments, order-1 and order-2 terms appear less frequently and with lower importance scores, while order-3 and order-4 terms dominate. In particular, the two most influential features are consistently \texttt{sig\_27} (order-4) and \texttt{sig\_13} (order~3), which together account for approximately $50$--$60\%$ of the total built-in importance. This suggests that higher-order iterated integrals capture the path properties most relevant for model discrimination. 

The key distinguishing characteristics of stochastic volatility models are precisely the fine-grained geometric properties of the trajectories---roughness, mean reversion, and volatility of volatility---which are naturally encoded in higher-order interactions between increments. It is therefore intuitive that order-3 and order-4 terms are more discriminative than lower-order terms, which mainly capture coarse features such as overall displacement and signed area.

\paragraph{Permutation Importance Analysis.}

We are aware that tree-based importance scores may partially favor higher-order terms due to scaling effects. To assess the robustness of our results, we complement the built-in XGBoost importance scores with scale-invariant measures based on permutation importance. The vectorized truncated signature of order~$4$ is represented as:
\[
\mathbf{vec}(S(\mathbf{X})^{\leq 4}_t)\, =\, \left(\,\texttt{sig\_0},\ \texttt{sig\_1},\ \texttt{sig\_2},\ 
\texttt{sig\_3},\ \dots,\ \texttt{sig\_29},\ \texttt{sig\_30}\,\right).
\]
In multi-index notation, where $0$ denotes the time component and $1$ the path component, the coordinates are:
\[
\mathbf{vec}(S(\mathbf{X})^{\leq 4}_t) = \left( 1,\, S^{0},\, S^{1},\, S^{00},\, S^{01},\, S^{10},\, S^{11},\, 
S^{000},\, \dots,\, S^{111},\, S^{0000},\, \dots,\, S^{1111} \right).
\]
Note that \texttt{sig\_2} corresponds to $S^1 = \int_0^t dX_s = X_t - X_0$, the net increment of the path.

XGBoost measures feature importance through the reduction in loss achieved when a feature is used in a tree split. Higher-order signature terms typically exhibit larger variability across paths than lower-order terms: while lower-order terms capture simple path properties such as net displacement and signed area, higher-order terms encode complex interactions between increments at multiple time scales, leading to greater variability across paths from different model classes. Features with larger variability naturally provide more splitting opportunities, which may artificially inflate their importance scores in tree-based methods.

\begin{figure}[!htbp]
\centering
\includegraphics[width=0.95\textwidth]{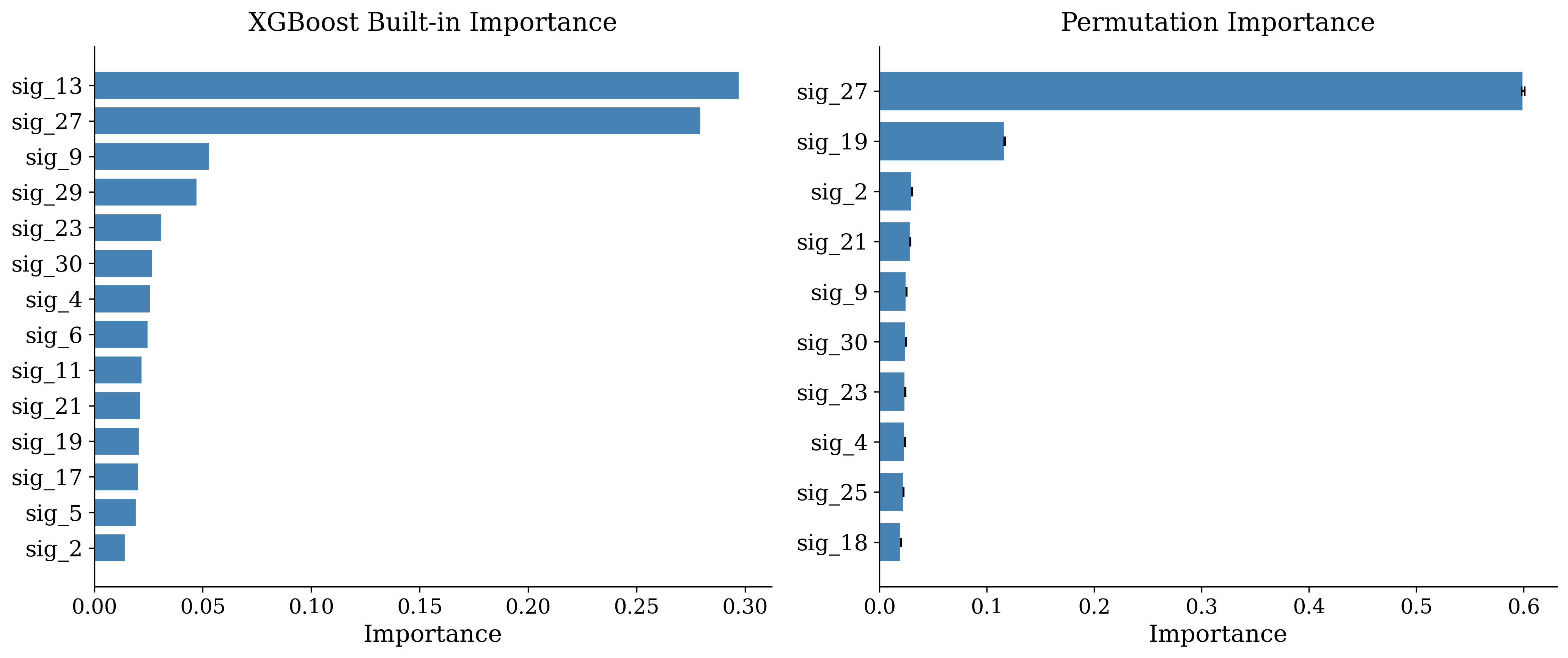}\\[0.8em]
\includegraphics[width=0.95\textwidth]{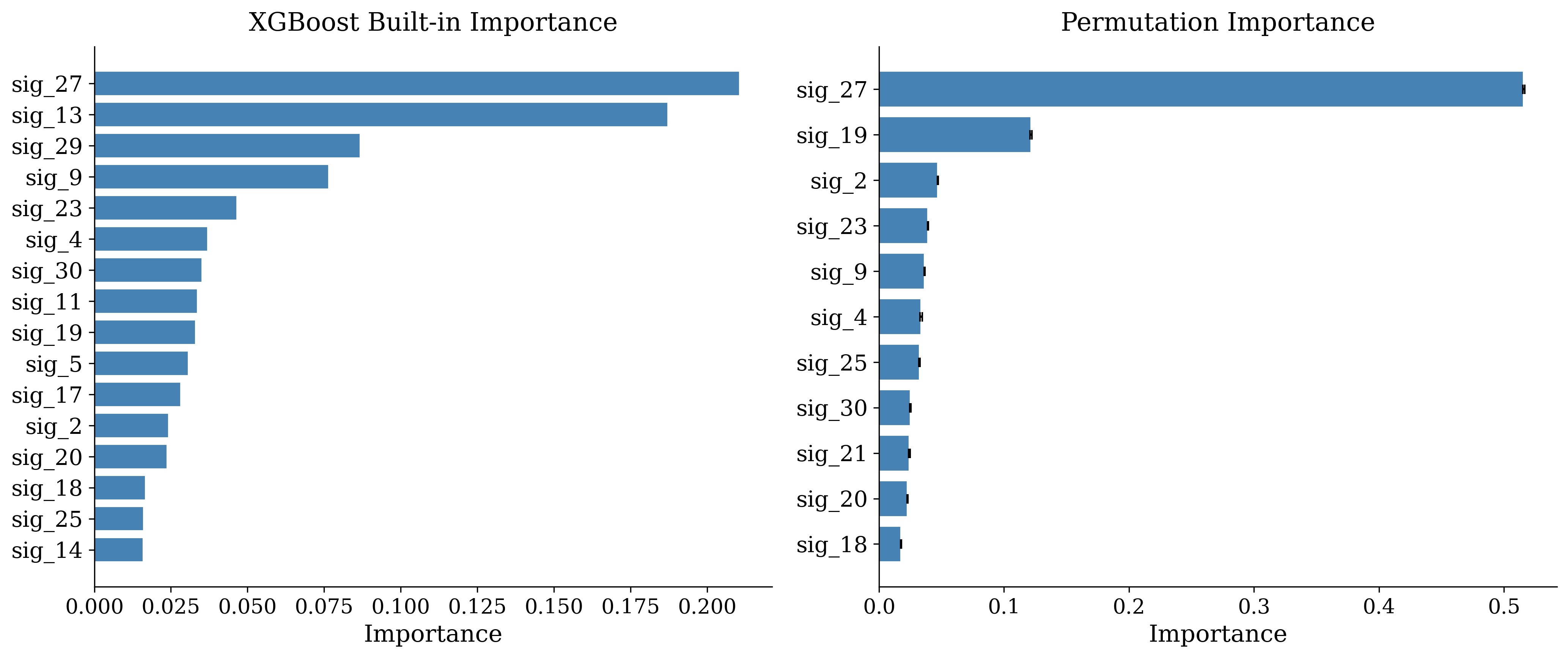}
\caption{XGBoost built-in importance (left) and permutation importance (right) for Experiments~\ref{exp_rnd2} (top) and~\ref{exp_rnd3} (bottom). Features are truncated at $90\%$ cumulative importance.}
\label{fig:perm_importance}
\end{figure}

To address this, we compute permutation importance (\cite{breiman01}), a scale-invariant alternative that measures the decrease in accuracy when each feature is randomly permuted. We report both importance measures for Experiments~\ref{exp_rnd2} and~\ref{exp_rnd3}, the two main experiments. The results are presented in Figure~\ref{fig:perm_importance}.

Across both experiments, the two importance measures agree on the dominant role of \texttt{sig\_27}, an order-4 signature element ($S^{1100}$), which accounts for approximately $50$--$60\%$ of the normalized permutation importance. A notable secondary feature under permutation importance is \texttt{sig\_19} ($S^{0100}$), which ranks second in both experiments with approximately $12\%$ of the importance, yet appears much lower in the built-in ranking. This divergence suggests that \texttt{sig\_19} is genuinely informative but suppressed in the built-in scores due to its smaller scale relative to higher-order terms.

The built-in and permutation importances also diverge on \texttt{sig\_13} ($S^{110}$): under the built-in measure it ranks among the top two features in both experiments, whereas under permutation importance it does not appear in the truncated list at all. This is consistent with the known tendency of tree-based importance measures to favor features with larger variance.

The feature \texttt{sig\_2} ($S^1 = X_t - X_0$) appears among the top features under permutation importance in both experiments, suggesting that the net displacement of the volatility path contributes to model discrimination alongside higher-order regularity information. Intuitively, mean-reverting models such as Heston and the Ornstein--Uhlenbeck process tend to produce paths with net displacement close to zero, whereas rough Bergomi paths may exhibit more sustained excursions.

We note that permutation importance is known to redistribute importance among correlated features, which signature terms of the same path inevitably are. The concentration of importance on \texttt{sig\_27} should therefore be interpreted with some caution, as correlated features may share discriminative information that is attributed entirely to the feature permuted first.

\subsection{Robustness with respect to signature order}

\begin{figure}[!htbp]
\centering
\includegraphics[width=0.44\textwidth]{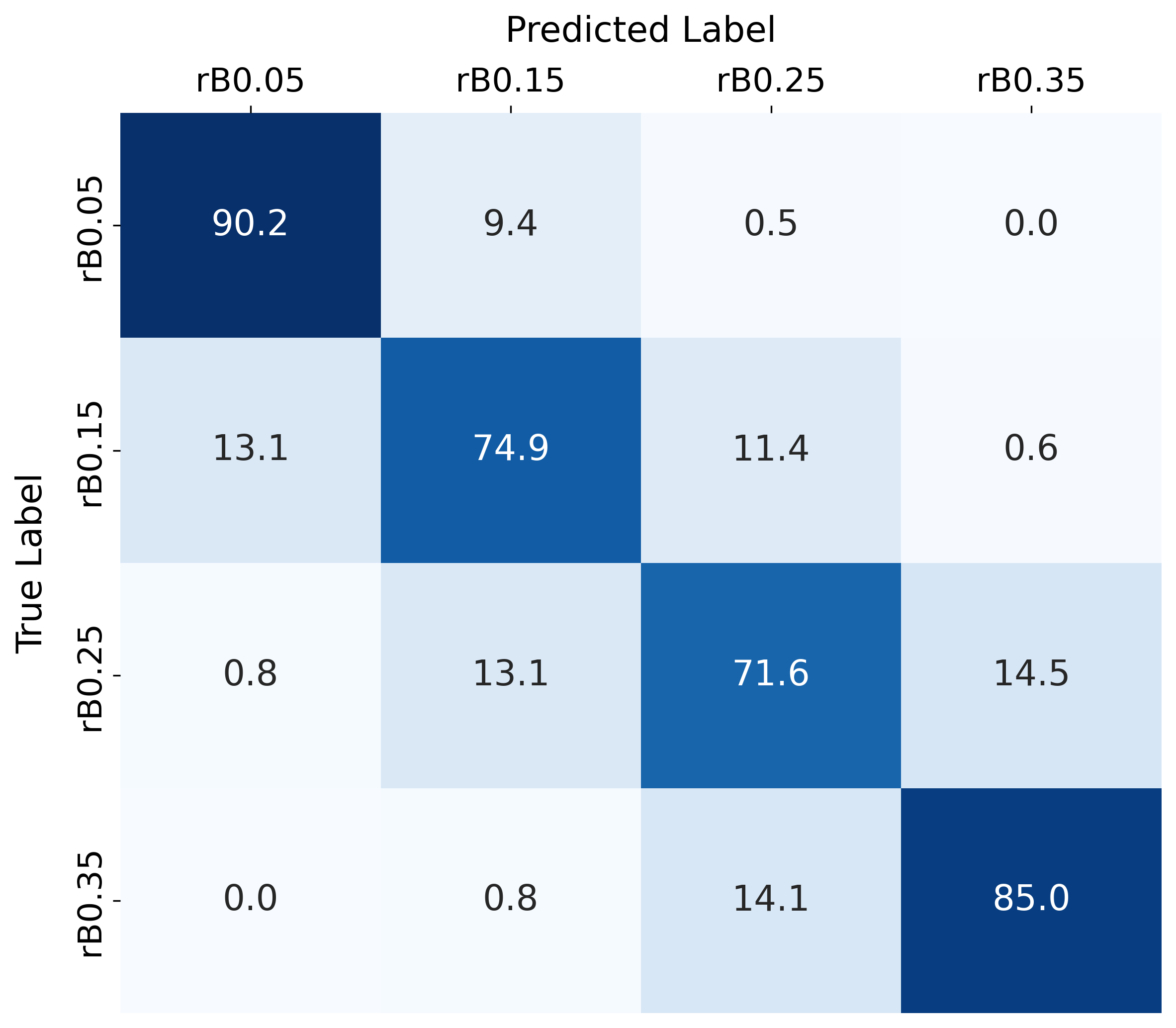}\hfill
\includegraphics[width=0.44\textwidth]{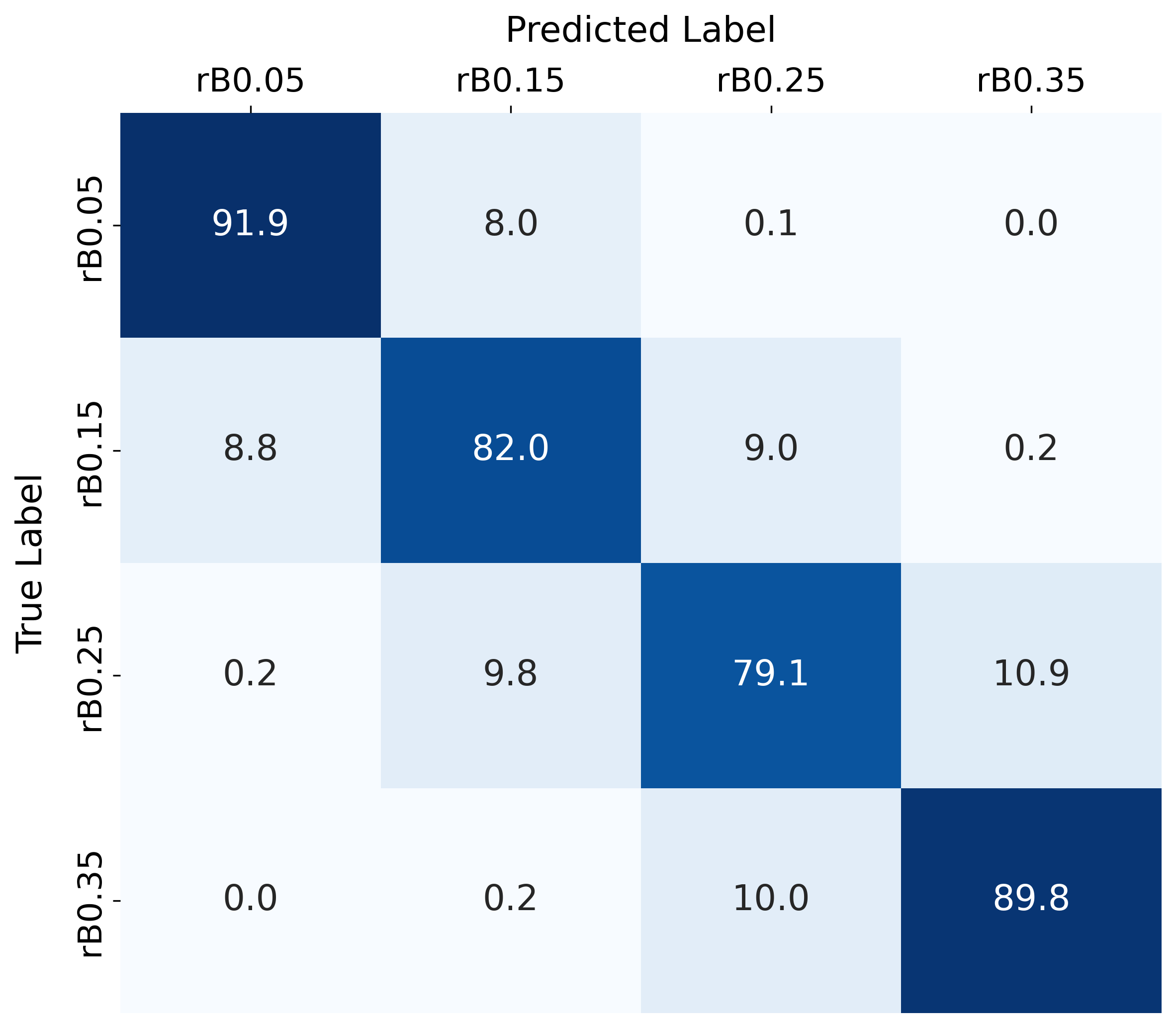}
\caption{Confusion matrices for Experiment~\ref{exp_rnd3} using signatures of order~$3$ (left) and order~$5$ (right) ($n_{\mathrm{test}} = 50{,}000$). Values are percentages.}
\label{fig:cmrnd3_S3S5}
\end{figure}

To further assess the role of the truncation level, we repeat Experiment~\ref{exp_rnd3} using signatures of order~$3$ and order~$5$. The corresponding confusion matrices are shown in Figure~\ref{fig:cmrnd3_S3S5}.

For order~$3$, although errors occur exclusively between models with adjacent Hurst parameters, their frequency is noticeably higher than in the order-$4$ case. This confirms that lower-order signatures lack some of the finer geometric detail needed to distinguish between models with very similar roughness.

For order~$5$, the vectorized signature satisfies $\mathbf{vec}(S(\mathbf{X})^{\leq 5}_t) \in \mathbb{R}^{63}$, and the 
results show a slight improvement in accuracy relative to order~$4$, with a reduced number of misclassified paths. The improvement, however, remains modest.

Although increasing the truncation level to order~$5$ yields a modest improvement in classification performance, it increases the computational time noticeably; see Table~\ref{tab:timings_rnd}. This suggests that the essential geometric characteristics distinguishing these models are already well captured at order~$4$, which offers a good balance between accuracy and efficiency.

\begin{table}[H]
\centering
\begin{tabular}{lccc}
\toprule
 & Total time & Signature computation & XGBoost training \\
\midrule
Experiment~\ref{exp_rnd1}           & 0:08:16 & 0:07:55 & 0:00:21 \\
Experiment~\ref{exp_rnd2}           & 0:07:23 & 0:07:05 & 0:00:18 \\
Experiment~\ref{exp_rnd3}           & 0:09:40 & 0:09:16 & 0:00:24 \\[4pt]
Experiment~\ref{exp_rnd3} (order 3) & 0:05:39 & 0:05:21 & 0:00:18 \\
Experiment~\ref{exp_rnd3} (order 5) & 0:15:27 & 0:15:03 & 0:00:24 \\
\bottomrule
\end{tabular}
\caption{Running times for experiments with randomly sampled model parameters.}
\label{tab:timings_rnd}
\end{table}

\subsection{Robustness with respect to time horizon.}

\begin{figure}[!htbp]
\centering
\includegraphics[width=0.44\textwidth]{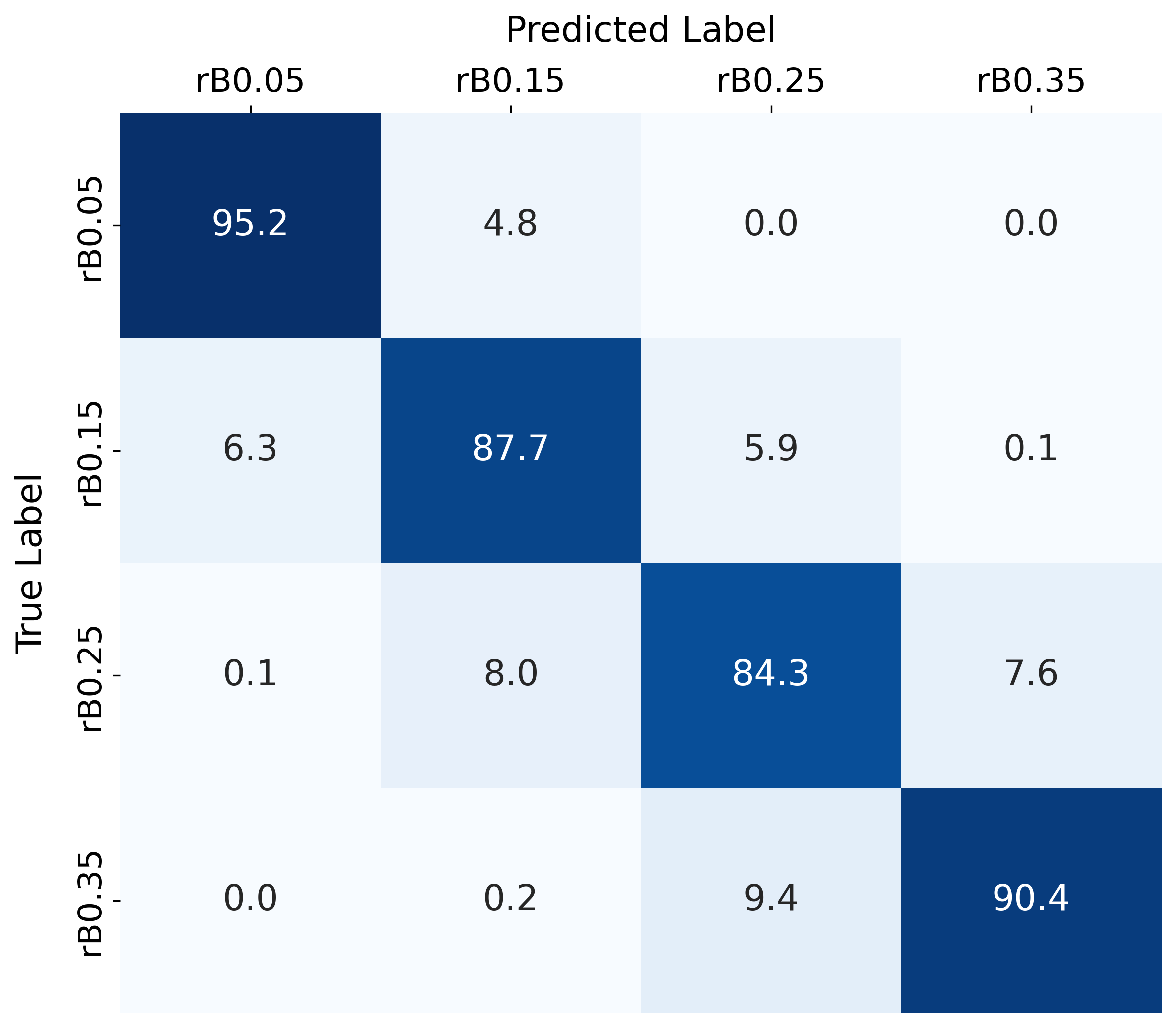}
\hfill
\includegraphics[width=0.44\textwidth]{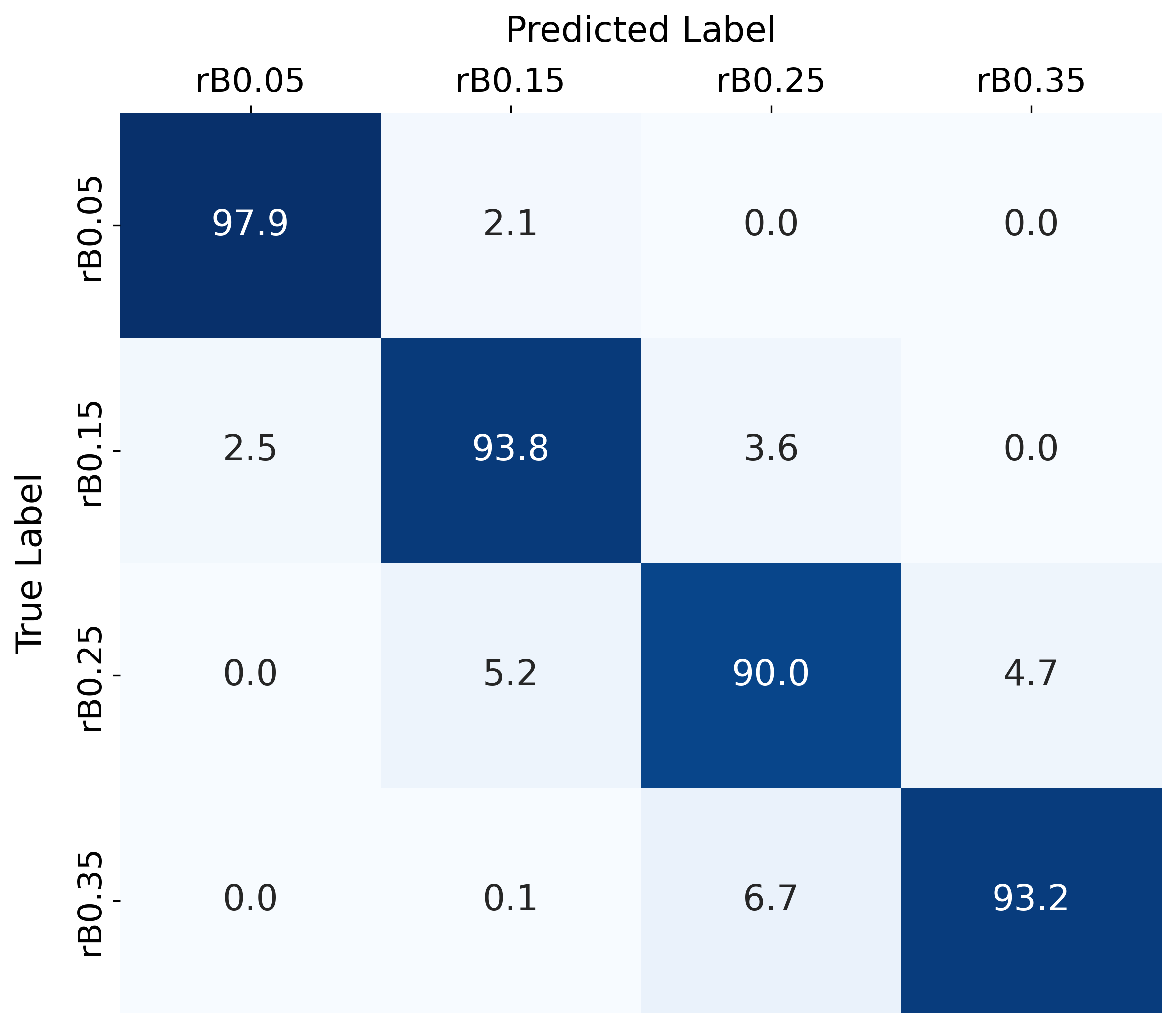}
\caption{Confusion matrices for Experiment~\ref{exp_rnd3} with $T=0.2$ (left) and $T=0.4$ (right). Values are percentages.}
\label{fig:cmrnd_T}
\end{figure}

As an additional robustness check, we investigate the effect of the time horizon $T$ on classification performance. All previous experiments were conducted with $T = 0.1$, a short-maturity for which rough volatility effects are most pronounced. We now repeat Experiment~3 with larger maturities, namely $T=0.2$ and $T=0.4$. The corresponding confusion matrices are shown in Figure~\ref{fig:cmrnd_T}.

We observe a noticeable improvement in classification accuracy for higher maturities, a behavior that is consistent with the fact that roughness is primarily a short-time phenomenon, and differences in the Hurst parameter are most difficult to detect at small maturities. As the time horizon increases, the cumulative effect of the dynamics becomes more pronounced, making the distinction between models easier for the classifier. Trajectories corresponding to longer maturities seem to provide additional information that enhances separability between volatility regimes.

\subsection{Comparison with neural networks}

We perform a robustness check by replacing the XGBoost classifier with a standard fully connected neural network applied to the same signature features. After standardizing the inputs, we train a network with three hidden layers of sizes $128$, $64$, and $32$ (a standard pyramidal architecture) using ReLU activations together with batch normalization and dropout regularization. The output layer uses a softmax activation for multi-class classification. The model is trained using the Adam optimizer with categorical cross-entropy loss. Since this is a one-off robustness check and the focus of the paper is on the XGBoost classifier, we do not perform hyperparameter tuning; a small validation split ($10\%$ of the training set) is used solely for early stopping to mitigate overfitting, and performance is reported on the test set.

\begin{figure}[!htbp]
\centering
\includegraphics[width=0.44\textwidth]{cm_rnd2.png}\hfill
\includegraphics[width=0.44\textwidth]{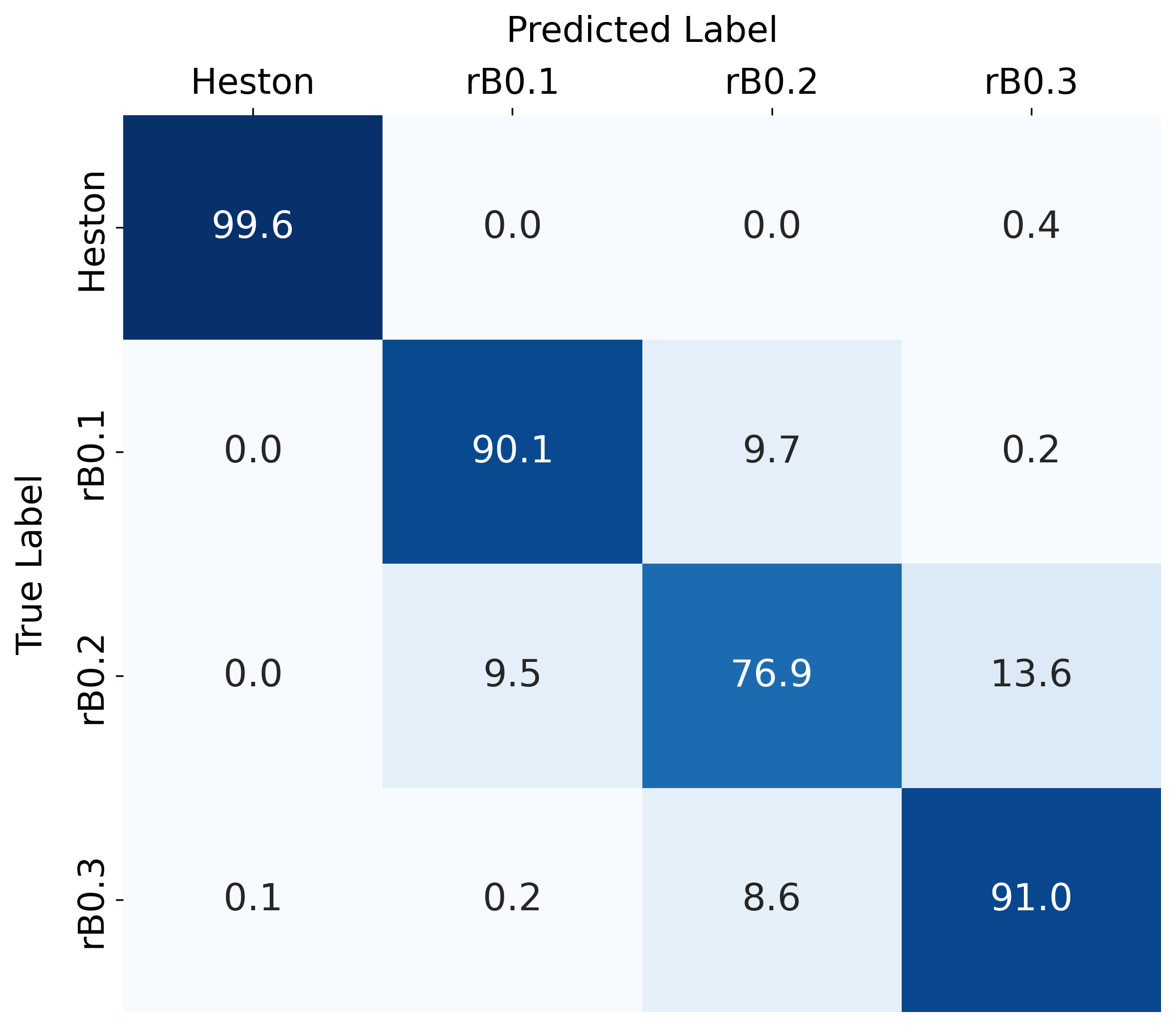}
\caption{Confusion matrices for Experiment~\ref{exp_rnd2} using XGBoost (left) and a neural network (right). Values are percentages 
($n_{\mathrm{test}} = 50{,}000$).}
\label{fig:cmrnd2_nn}
\end{figure}

We compare with Experiment~\ref{exp_rnd2}, which involves the classification of a Heston model and three rough Bergomi models with 
different Hurst parameters. The confusion matrices for XGBoost and the neural network are shown side by side in Figure~\ref{fig:cmrnd2_nn}.

The two classifiers achieve broadly comparable accuracy on this experiment. The Heston class is identified with $100\%$ accuracy by XGBoost and $99.6\%$ by the neural network. Misclassifications occur exclusively between adjacent rough Bergomi classes in both cases, with the neural network showing slightly higher confusion for rB0.2 and slightly lower confusion for rB0.3. Overall, the differences in classification accuracy are modest, confirming that the signature features carry sufficient discriminative information regardless of the classifier used.

The neural network is, however, significantly more expensive to train. While XGBoost required only $18$ seconds, the neural network training with $40$ epochs took more than $5$ minutes under the same conditions. In addition, the neural network does not provide a direct measure of feature importance, in contrast to the tree-based approach.

Overall, these results confirm that gradient boosting methods are well suited to this problem. In the present setting, where the input data are structured feature vectors derived from signatures, XGBoost provides comparable or superior accuracy, faster training, and greater interpretability than the neural network alternative.

\subsection{A robustness check for Heston and Ornstein--Uhlenbeck model classes} \label{sec:OUH_robustness}

\begin{figure}[!htbp]
\centering
\includegraphics[width=0.44\textwidth]{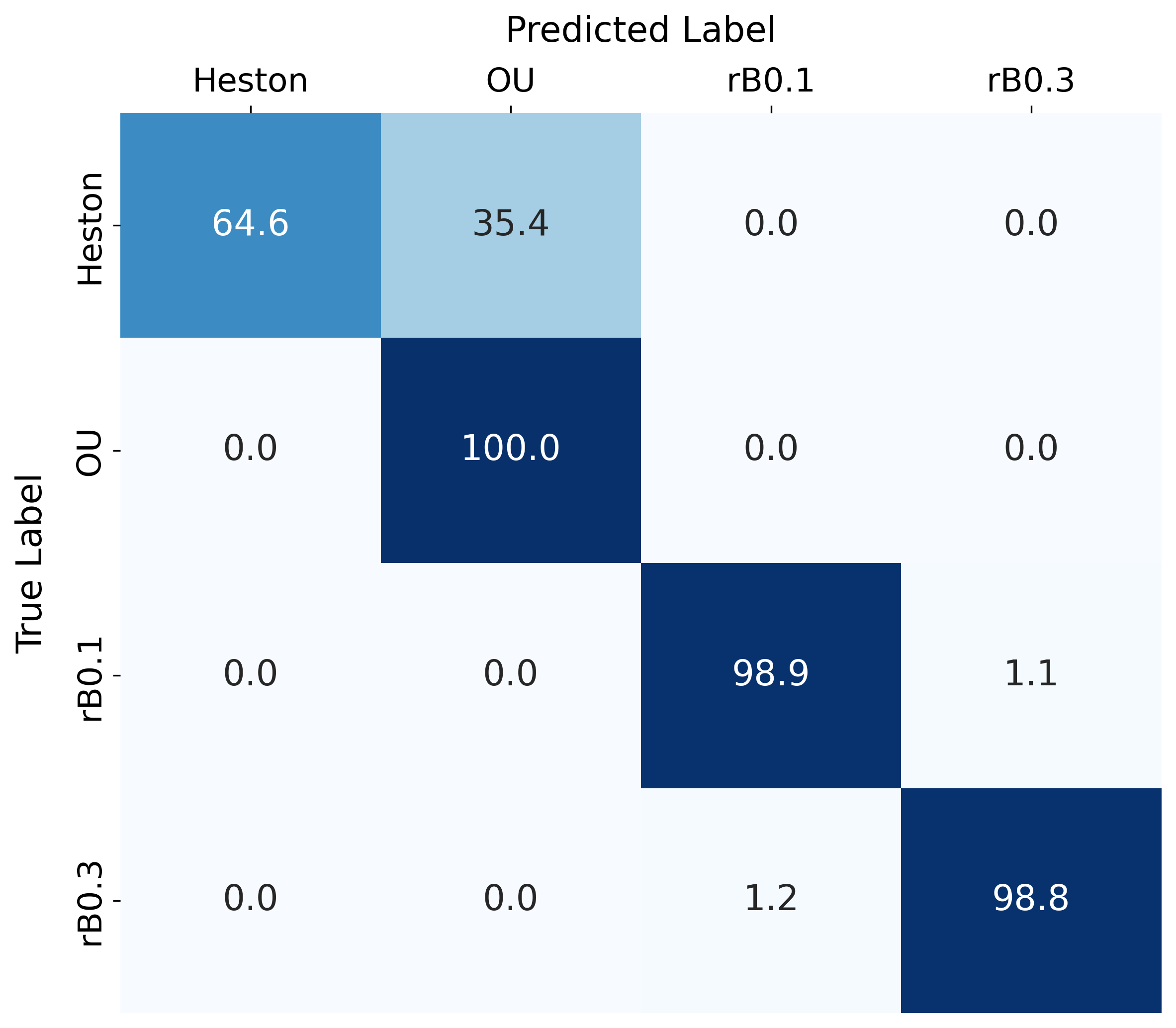}
\caption{Confusion matrix for Experiment~\ref{exp_rnd1} with identical parameter distributions for the Heston and Ornstein--Uhlenbeck processes ($n_{\mathrm{test}} = 50{,}000$). Values are percentages.}
\label{fig:HOU_robust}
\end{figure}

To further stress-test the classifier, we consider an experiment in which the Heston and Ornstein--Uhlenbeck processes are simulated using identical parameter distributions.\footnote{Both processes share identical distributions for the mean-reversion parameters $\kappa$ and $\theta$. The diffusion coefficients are sampled from comparable ranges, the Heston volatility-of-volatility parameter $\nu$ being additionally constrained by the Feller condition.} In this setting, any successful classification must be attributed entirely to the structural difference in their diffusion terms: $\nu\sqrt{X_t}$ in the Heston model versus the constant coefficient $\sigma$ in the Ornstein--Uhlenbeck process. The results are reported in Figure~\ref{fig:HOU_robust}.

As expected, the classification of the rough Bergomi classes remains essentially unaffected: both rB0.1 and rB0.3 are identified with $98.9\%$ accuracy, and no confusion occurs between the rough and non-rough classes. This confirms that the distinction between rough and Markovian volatility dynamics is robust to parameter uncertainty.

The main challenge arises in distinguishing Heston from Ornstein--Uhlenbeck. When the two processes share the same parameter distributions, the Heston class is correctly identified only $64.6\%$ of the time, with $35.4\%$ of Heston paths misclassified as Ornstein--Uhlenbeck. Conversely, the Ornstein--Uhlenbeck class is identified with perfect accuracy, suggesting that the classifier has learned a decision boundary biased toward the Ornstein--Uhlenbeck class when the two processes become geometrically similar.

This asymmetry is noteworthy. When model classes differ primarily through path regularity (as in rough Bergomi models with different Hurst parameters) the trajectories remain globally distinct, and the signature captures these differences through higher-order interactions. By contrast, when two processes can become locally equivalent, as in the present Heston/OU setting, the available geometric information is intrinsically limited. In this case, the challenge does not arise from a weakness of the signature representation, but from the genuine similarity between the underlying stochastic dynamics.

\begin{figure}[!htbp]
\centering
\includegraphics[width=0.99\textwidth]{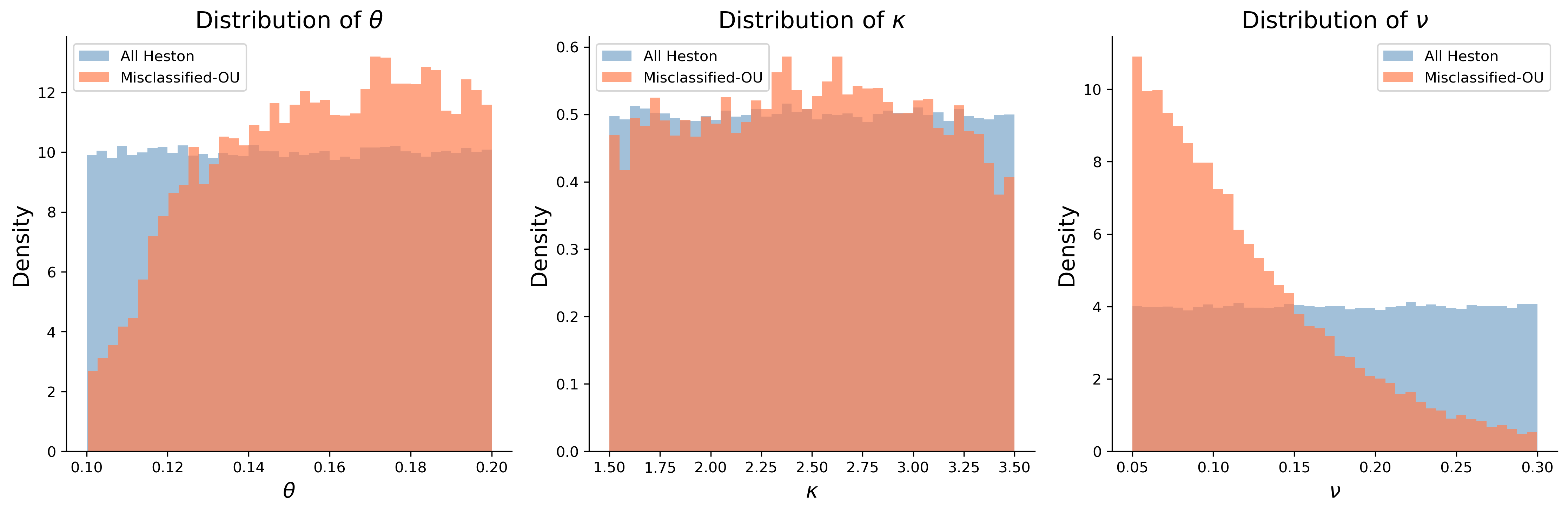}
\caption{Empirical distributions of the Heston parameters $\theta$, $\kappa$, and $\nu$ for all simulated Heston paths and for the subset misclassified as Ornstein--Uhlenbeck.}
\label{fig:heston_misclassified}
\end{figure}

To better understand the source of the misclassifications, we examine the distribution of model parameters for the Heston paths incorrectly classified as Ornstein--Uhlenbeck. Figure~\ref{fig:heston_misclassified} compares the empirical distributions of the Heston parameters across all simulated paths (blue) with the corresponding distributions restricted to the subset of paths misclassified as Ornstein--Uhlenbeck (orange).

As can be seen from the middle histogram, the mean-reversion speed $\kappa$ exhibits no systematic difference between correctly classified and misclassified trajectories, confirming that it plays no role in the misidentification. In the left histogram, the mean value of $\theta$ is essentially identical across the two groups ($0.157$ for misclassified vs $0.150$ for all Heston paths), indicating that the long-run mean plays no meaningful role. By contrast, data from the right histogram reveal a pronounced shift in the distribution of $\nu$: misclassified paths have a substantially lower mean volatility of volatility ($0.120$ vs $0.175$ for all Heston paths). This identifies $\nu$ as the primary driver of the misclassification.

The mechanism is intuitive. When $\nu$ is small, the diffusion term $\nu\sqrt{X_t}$ of the Heston variance process contributes little to the path dynamics, which are then dominated by the drift term $\kappa(\theta - X_t)\,dt$. In this regime, the Heston process behaves almost like a deterministic mean-reverting process perturbed by a small diffusion, making it geometrically indistinguishable from an Ornstein--Uhlenbeck process with the same drift parameters and a similarly small constant diffusion coefficient $\sigma$. The state-dependence of $\nu\sqrt{X_t}$ simply does not manifest when $\nu$ is too small for the signature to detect.

\begin{figure}[!htbp]
\centering
\includegraphics[width=0.44\textwidth]{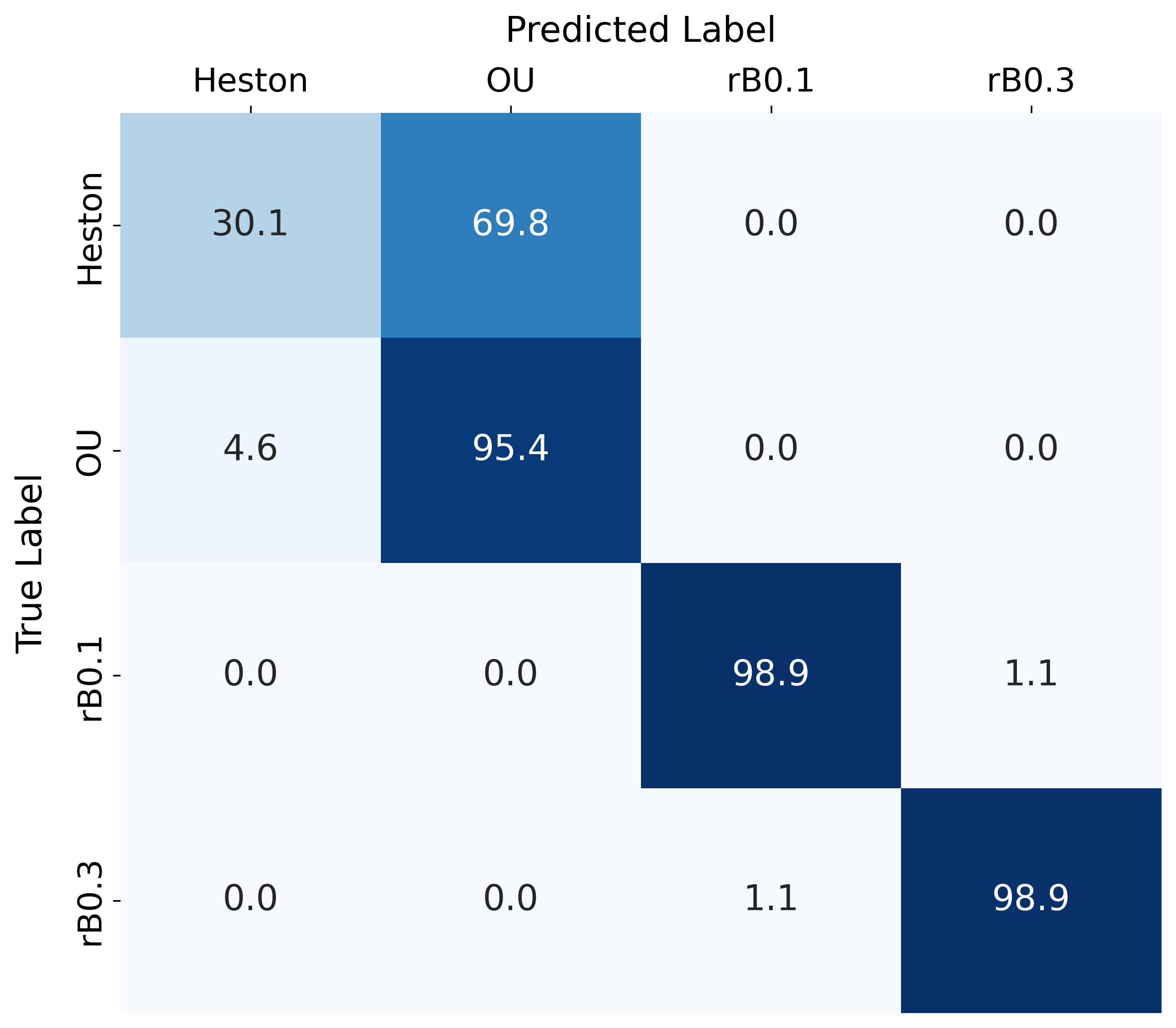}\hfill
\includegraphics[width=0.44\textwidth]{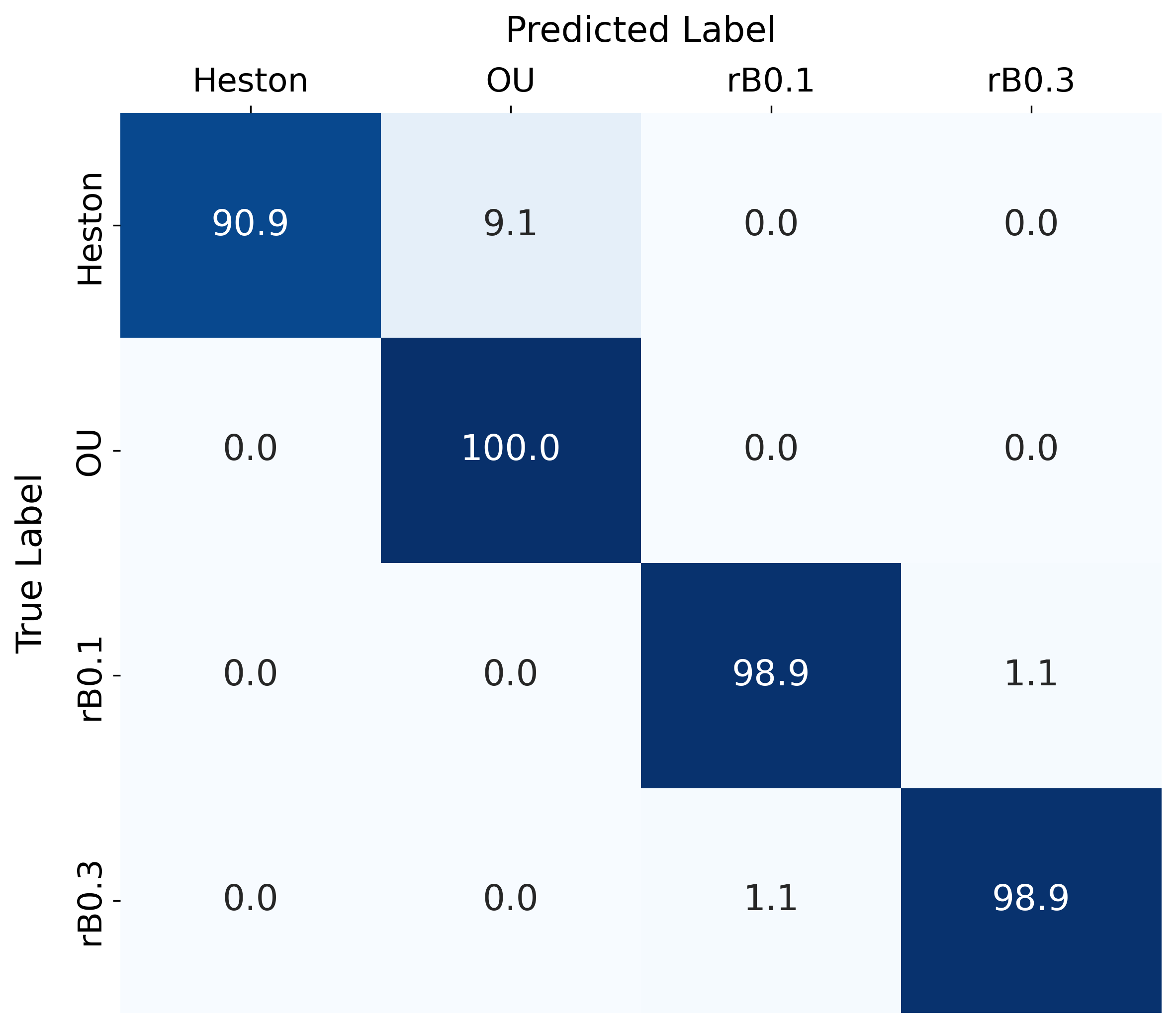}
\caption{Confusion matrices for the Heston/OU experiment with low $\nu \sim U(0.01, 0.10)$ (left) and high $\nu = 0.28$ fixed (right) ($n_{\mathrm{test}} = 50{,}000$). Values are percentages.}
\label{fig:cm_fixednu}
\end{figure}

To confirm that $\nu$ is the primary driver of the misclassification, we repeat the experiment twice with $\nu$ fixed at a low value ($\nu \sim U(0.01, 0.10)$) and at a high value ($\nu = 0.28$), keeping all other parameters unchanged. The resulting confusion matrices are displayed in Figure~\ref{fig:cm_fixednu}.

The results show that when $\nu$ is small, the diffusion component $\nu\sqrt{X_t}$ contributes little to the path dynamics, making the Heston process nearly indistinguishable from an Ornstein--Uhlenbeck process with the same drift parameters: Heston accuracy falls to $30.1\%$, with $69.8\%$ of paths misclassified as Ornstein--Uhlenbeck. Notably, the asymmetry partially breaks down in this regime, as the Ornstein--Uhlenbeck accuracy also deteriorates to $95.4\%$, with $4.6\%$ of OU paths misclassified as Heston, suggesting that when $\nu$ is very small the two processes become genuinely difficult to distinguish in both directions.

By contrast, when $\nu = 0.28$, the state-dependence of the diffusion term is geometrically pronounced and detectable from the signature: Heston accuracy recovers to $90.9\%$ and Ornstein--Uhlenbeck remains at $100\%$. The rough Bergomi classes are unaffected in both experiments ($98.9\%$ in each case), confirming that $\nu$ influences only the Heston/OU boundary and has no bearing on the classification of rough volatility dynamics. The residual $9.1\%$ misclassification rate in the high-$\nu$ case suggests that some geometric similarity between the two processes persists over short time horizons, even when the state dependence of the Heston diffusion is pronounced.

\begin{figure}[!htbp]
\centering
\includegraphics[width=0.95\textwidth]{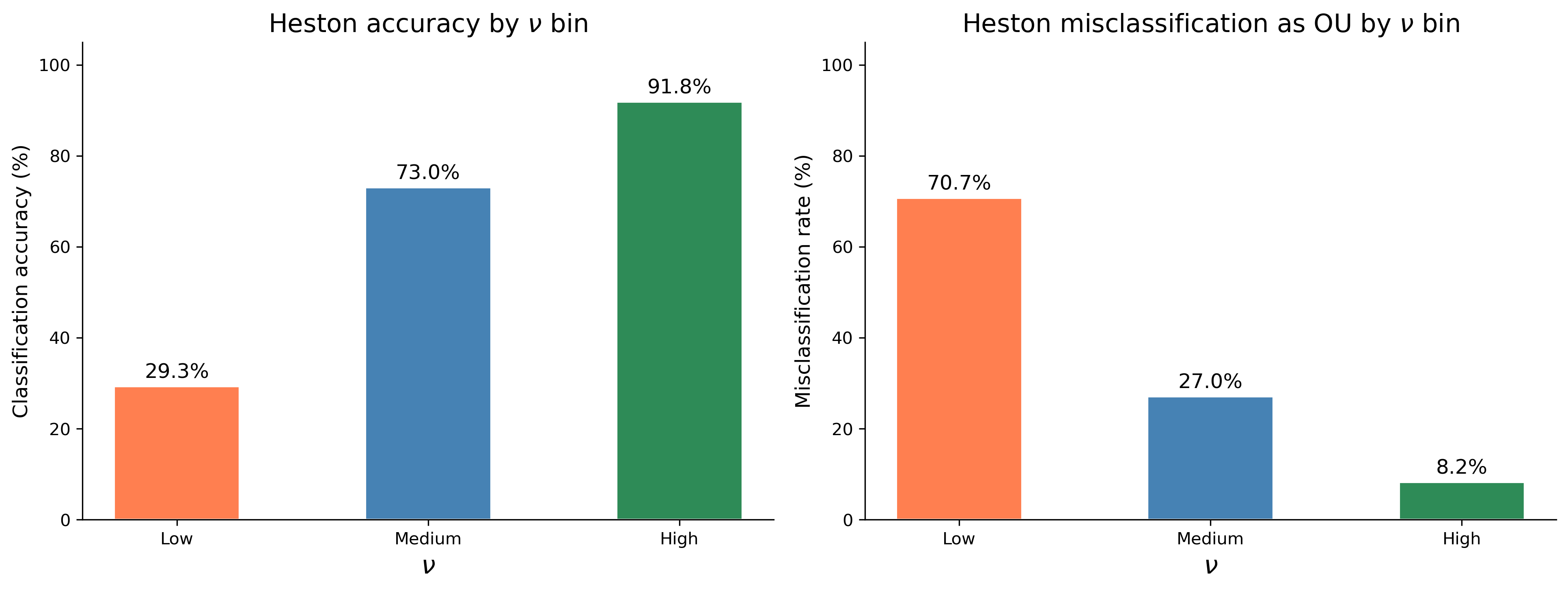}
\caption{Heston classification accuracy (left) and misclassification rate into the OU class (right) across three equal-sized empirical bins of the volatility of volatility parameter $\nu$.}
\label{fig:acc_nu}
\end{figure}

To further confirm this interpretation, we partition the Heston test paths into three groups according to the value of $\nu$, using the $1/3$ and $2/3$ quantiles of the empirical distribution as bin boundaries (see Appendix~\ref{app:nu_bins} for the exact boundaries and statistics). For each group, we compute the classification accuracy and the misclassification rate into the Ornstein--Uhlenbeck class. The results are shown in Figure~\ref{fig:acc_nu}, which confirms the theoretical interpretation. 

The misclassification rate decreases monotonically from $70.7\%$ in the low regime to $27.0\%$ in the medium regime and $8.2\%$ in the high regime, while accuracy increases from $29.3\%$ to $73.0\%$ and $91.8\%$. This monotone relationship between $\nu$ and classification performance confirms that the volatility of volatility is the key parameter governing the difficulty of distinguishing Heston from Ornstein--Uhlenbeck, and that the observed misclassifications arise from genuine geometric similarities between the two processes rather than from a limitation of the signature representation.

\subsection{Sample size and computational trade-offs} \label{sec:sample_size}

All experiments presented in this paper were conducted using $250{,}000$ simulated paths per model class. This large sample size ensures stable training and provides strong empirical support for the classification results. The trade-off, however, is an increase in computational time.

To assess the sensitivity of the methodology to the sample size, we repeated the main experiments using only $50{,}000$ paths per model class (that is, $200{,}000$ paths per experiment instead of one million). The results are summarized in Table~\ref{tab:sample_size}.

\begin{table}[!htbp]
\centering
\begin{tabular}{lcccc}
\toprule
 & Time (50K) & Accuracy (50K) & Time (250K) & Accuracy (250K) \\
\midrule
Experiment~\ref{exp_rnd1} & 0:00:59 & 0.9846 & 0:08:16 & 0.9863 \\
Experiment~\ref{exp_rnd2} & 0:01:28 & 0.8928 & 0:07:23 & 0.8956 \\
Experiment~\ref{exp_rnd3} & 0:02:58 & 0.8364 & 0:09:40 & 0.8391 \\
\bottomrule
\end{tabular}
\caption{Running times and test accuracies for different sample sizes.}
\label{tab:sample_size}
\end{table}

The results indicate that XGBoost maintains high classification accuracy even with substantially smaller training datasets. Although a slight reduction in performance is observed, the loss in accuracy remains modest relative to the significant reduction in computational time. The proposed framework therefore allows for a practical trade-off between computational speed and predictive accuracy, depending on the requirements of the application.

\section{Conclusion} \label{sec:conclusion}

In this paper, we have proposed a signature-based framework for identifying volatility dynamics directly from path geometry, without relying on parametric calibration or prior knowledge of the underlying model parameters.

In the first part of the paper, each model class is simulated with fixed parameter values. These experiments serve as a proof of concept showing that signatures distinguish path geometries with extremely high accuracy. In simple settings, where the models differ structurally, classification is essentially perfect. As the task becomes more challenging, with models differing only through the degree of roughness, the method continues to perform robustly, with errors occurring primarily between trajectories generated with nearby values of the Hurst parameter.

In the second part of the paper, we address the more demanding problem of \emph{model-class identification} under parameter uncertainty. Here, the classifier no longer distinguishes trajectories generated from fixed parameter configurations, but rather entire model classes whose parameters are sampled randomly within prescribed ranges. Although classification accuracy naturally decreases in this setting, the results show that it remains high overall. In particular, the experiments reveal a fundamental distinction between two types of classification difficulty. Rough volatility models with nearby Hurst parameters become difficult to distinguish because their trajectories exhibit similar levels of roughness, whereas processes such as Heston and Ornstein--Uhlenbeck become increasingly difficult to separate as their diffusion structures become locally similar under low volatility of volatility regimes.

The robustness analysis of the Heston/Ornstein--Uhlenbeck setting provides further insight into this phenomenon. Misclassification occurs predominantly when the volatility of volatility parameter $\nu$ is small: in this regime, the state-dependent diffusion term 
$\nu\sqrt{X_t}$ of the Heston model contributes little to the path dynamics, making the process geometrically indistinguishable from an Ornstein--Uhlenbeck process with the same drift parameters. This is confirmed by a monotone relationship between $\nu$ and classification accuracy: when the Heston test paths are partitioned into three equal-sized groups according to $\nu$, the misclassification rate decreases from $70.7\%$ in the low-$\nu$ regime to $27.0\%$ and $8.2\%$ in the medium- and high-$\nu$ regimes, respectively. These results suggest that the remaining classification errors reflect intrinsic geometric similarities between the underlying stochastic dynamics, rather than limitations of the signature representation itself.

The analysis of feature importance provides additional insight into the mechanism underlying these results. Across all experiments, the most relevant features are consistently drawn from levels~$3$ and~$4$ of the signature, confirming that classification relies primarily on higher-order interactions between increments rather than on marginal path properties alone. Increasing the truncation level beyond order~4 yields only modest gains, indicating that the essential geometric characteristics distinguishing these models are already well captured by signatures truncated at order~4.

From a computational perspective, the proposed methodology remains practical. The main computational cost lies in path simulation and signature computation; XGBoost training is comparatively inexpensive and completes in seconds even for large datasets. Vectorized implementations and GPU acceleration keep the total runtime within minutes, and the method maintains high accuracy even with substantially fewer simulated trajectories, allowing for a flexible trade-off between computational cost and predictive precision.

Finally, the robustness analysis with respect to the time horizon highlights that rough volatility effects are most subtle at short maturities, where classification is most challenging. As the maturity increases, differences in path geometry become more pronounced and classification accuracy improves. In this sense, the short-maturity experiments considered throughout the paper may be regarded as a conservative scenario.

Taken together, these results support the view that volatility dynamics can be effectively identified through the geometry of their sample paths, and suggest that signature features may provide a robust alternative to classical Hurst parameter estimation methods \citep{GatheralJaiRosen18, bennedsen22} and to neural network approaches \citep{stone20} within a general model-identification framework that does not require task-specific architectures or explicit parameter estimation.

Our experiments suggest that signatures distinguish stochastic dynamics not merely through local statistical properties, but through the global geometric structure of the trajectories. The resulting framework is flexible, interpretable, and robust under parameter uncertainty. More generally, path signatures may serve as a general-purpose feature map for model identification in settings beyond stochastic volatility, wherever the relevant information is encoded in the geometry of observed paths.

A natural limitation of the present study is that the analysis is conducted entirely on simulated data. Extending the methodology to observed realized volatility series, and assessing whether the classifier trained on simulated paths generalizes to market data, is an important direction for future work.

\vspace{1cm}

\appendix
\section*{\Large Appendix}
\section{Volatility of Volatility Bins for the Heston--OU Analysis} \label{app:nu_bins}

Table~\ref{tab:nu_bins} reports the quantile boundaries and summary statistics for the analysis of $\nu$ in Section~\ref{sec:OUH_robustness}. The bin boundaries are defined as the~$1/3$ and~$2/3$ percentiles of the empirical distribution of $\nu$ across all Heston paths, giving approximately equal group sizes.

\begin{table}[H]
\centering
\begin{tabular}{lcccc}
\toprule
Bin & $N$ paths & Accuracy & Misclassification rate & Mean $\nu$ \\
\midrule
Low\phantom{m}   ($\nu \leq 0.1328$)  & 16{,}625 & $29.3\%$ & $70.7\%$ & $0.0915$ \\
Medium ($0.1328 < \nu \leq 0.2179$)     & 16{,}934 & $73.0\%$ & $27.0\%$ & $0.1750$ \\
High\phantom{m}  ($\nu > 0.2179$)      & 16{,}441 & $91.8\%$ & $8.2\%$ & $0.2591$ \\
\bottomrule
\end{tabular}
\caption{Accuracy and misclassification rate for Heston paths partitioned by the $1/3$ and $2/3$ quantiles of the empirical distribution of $\nu$ ($0.133$ and $0.218$, respectively).}
\label{tab:nu_bins}
\end{table}

\vspace{1em}

\bibliographystyle{apalike}
\bibliography{references.bib}

\end{document}